%% file: main.tex
\documentclass[10pt,conference]{IEEEtran}
\IEEEoverridecommandlockouts
\usepackage{cite}
\usepackage{amsmath,amssymb,amsfonts}
\usepackage{algorithmic}
\usepackage{graphicx}
\usepackage{textcomp}
\usepackage{xcolor}
\usepackage{subcaption}
\usepackage{hyperref}
\usepackage{soul}
\usepackage{bbm}
\usepackage[ruled,vlined,linesnumbered]{algorithm2e}
\usepackage{booktabs}
\usepackage{amsmath}
\newcommand{\uixpose}[1]{\textsc{UIXpose}}
\usepackage{tikz}
\usepackage{pgfplots}
\pgfplotsset{compat=1.18}

\usepackage{booktabs,tabularx,makecell}

\def\BibTeX{{\rm B\kern-.05em{\sc i\kern-.025em b}\kern-.08em
    T\kern-.1667em\lower.7ex\hbox{E}\kern-.125emX}}
\begin{document}

% \title{UIXpose: Towards Effective Mobile Malware Detection via Automated UI Exploration and Deep System Inspection}

% \title{UIXPOSE: Aligning Screen Intent with Runtime Behaviour for Mobile Malware Detection}

\title{
% \uixpose{}: Intention–Behaviour Alignment with Runtime Semantics for Mobile Malware Detection
\uixpose{}: Mobile Malware Detection via Intention–Behaviour Discrepancy Analysis
}

\author{
\IEEEauthorblockN{
Amirmohammad Pasdar,
Toby Murray,
Van-Thuan Pham
}
\IEEEauthorblockA{
\textit{School of Computing and Information Systems} \\
\textit{The University of Melbourne}, Melbourne, Australia \\
\{apasdar, toby.murray, thuan.pham\}@unimelb.edu.au
}
}
% toby.murray@unimelb.edu.au, thuan.pham@unimelb.edu.au
% \author{\IEEEauthorblockN{Anonymous Authors}}

% \author{\IEEEauthorblockN{Amirmohammad Pasdar}
% \IEEEauthorblockA{\textit{School of CIS} \\
% \textit{The University of Melbourne}\\
% % City, Country \\
% apasdar@unimelb.edu.au}
% \and
% \IEEEauthorblockN{Toby Murray}
% \IEEEauthorblockA{\textit{School of CIS} \\
% \textit{The University of Melbourne}\\
% % City, Country \\
% toby.murray@unimelb.edu.au}
% \and
% \IEEEauthorblockN{Van-Thuan Pham}
% \IEEEauthorblockA{\textit{School of CIS} \\
% \textit{The University of Melbourne}\\
% % City, Country \\
% thuan.pham@unimelb.edu.au
% }
% }

\maketitle

\begin{abstract}
We introduce \uixpose{}, a source-code-agnostic framework that operates on both compiled and open-source apps. This framework applies Intention–Behaviour Alignment (IBA) to mobile malware analysis, aligning UI-inferred intent with runtime semantics. Previous work either infers intent statically, e.g., permission-centric, or widget-level or monitors coarse dynamic signals (endpoints, partial resource usage) that miss content and context. \uixpose{} infers an intent vector from each screen using vision–language models and knowledge structures, and combines decoded network payloads, heap/memory signals, and resource utilisation traces into a behaviour vector. Their alignment, calculated at runtime, can both detect misbehaviour and highlight exploration of behaviourally rich paths. In three real-world case studies, \uixpose{} reveals covert exfiltration and hidden background activity that evade metadata-only baselines, demonstrating how IBA improves dynamic detection.
\end{abstract}

% hybrid detection is because the automated GUI tools rely on static analysis, so LLMS can help use such context and IBA for better exploration. This is the future work.
% Any parts of guiding the exploration with misalignment are a strong work for extension. 

\begin{IEEEkeywords}
Mobile Malware Detection, Intention–Behaviour Alignment, automated GUI testing, artificial intelligence
\end{IEEEkeywords}

\input{sections/intro}
\input{sections/motivation}
\input{sections/approach}
\input{sections/evaluation}
\input{sections/discussion}
\input{sections/conclusion}

\bibliographystyle{ieeetr}
\bibliography{references}

\end{document}

%% file: sections/intro.tex
\section{Introduction}\label{sec:intro}

Mobile applications (a.k.a. mobile apps) have become integral to modern life, facilitating everything from personal communication to business operations. As their use expands, so does their appeal to attackers seeking to steal personal data, sensitive corporate information, or even classified government secrets. Despite the seriousness of these threats, many malicious apps still evade detection, even in popular app stores, due to limitations in current malware detection methods \cite{slopadds}.

Malware detection methods are typically categorised into static \cite{pendlebury2019tesseract, idrees2017pindroid, wu2023droidrl, zhu2023android, ksibi2024efficient, li2017static}, dynamic \cite{wuchner2017leveraging, ribeiro2020hidroid, bhatia2017malware, keyes2021entroplyzer}, and hybrid approaches \cite{saracino2016madam, abawajy2017iterative, tong2017hybrid, arora2018ntpdroid}. Static analysis, including emerging techniques powered by large language models (LLMs) \cite{wang2025contemporary,zhao2025apppoet}, can efficiently scale to analyse large numbers of apps. However, it often suffers from high rates of false positives and false negatives. Since it relies heavily on code pattern matching across both control and data flows, static analysis may struggle to detect new malware that employs obfuscation or polymorphism. Very specific patterns can lead to false negatives, while overly broad patterns can lead to false positives.

Dynamic analysis observes behaviour during execution, but its success depends on reaching the correct paths and gathering meaningful signals. In real-world scenarios, even the most advanced GUI-driven exploration typically covers only about 20–30\% of activity in actual apps, leaving much behaviour unobserved  \cite{akinotcho2025mobile}. Additionally, client-side protections, such as certificate pinning, often prevent man-in-the-middle inspection of TLS traffic, thereby forcing systems to rely on broad indicators, such as destination and flow-level statistics \cite{pradeep2022comparative}. These wide signals can be easily bypassed, as adversaries quickly change endpoints using techniques such as fast-flux or domain generation algorithms (DGAs), and increasingly use encryption to conceal payload content, thereby reducing the effectiveness of blacklists or metadata-based detection \cite{ipflux}. Finally, hybrid approaches aim to capitalise on the strengths of both static and dynamic analysis. However, in practice, they still inherit many of the core limitations of both methods, which restrict their overall effectiveness. These analytical blind spots are especially problematic because they lead to behaviours that neither automated systems nor end-users can easily notice, emphasising a fundamental imbalance between what apps actually do and what users can observe.

% From a user’s perspective, this discrepancy is significant, as the screen effectively communicates the app's intended functions; yet, sensitive behaviours may occur outside the user's view, where visibility is limited. Recent work begins to connect what users see to what apps do across several fronts. On the UI side, researchers detect deceptive or misleading designs (dark patterns) directly from screenshots \cite{mansur2023aidui} and learn robust visual similarity metrics for Android screens \cite{li2024uihash}, even from a permission-alignment perspective via static analysis \cite{xi2019deepintent, yang2022describectx}, thereby improving how we read interfaces. On the behavioural side, privacy-compliance systems align claimed disclosures with actual flows, checking flow-to-policy consistency at scale \cite{tan2023ptpdroid}, or combining static flows with dynamic traffic to catch non-compliant transmissions \cite{fan2024giving}. In exploration, LLM-driven agents now plan over UI state to drive interactions \cite{liu2024make, feng2024prompting}, and a runtime framework \cite{ran2024guardian} offloads navigation to navigate screens better. Collectively, these advances strengthen UI understanding, compliance checking, and UI-guided exploration; however, they fall short of achieving a runtime, screen-level alignment between inferred user-visible intent and observed execution-time behaviour (including runtime semantics) that can actively steer exploration and flag misalignment as a first-class signal.

From a user’s perspective, this discrepancy is significant as the screen effectively communicates the app's intended functions, yet sensitive behaviours may occur outside the user's view, where visibility is limited. Recent work begins to connect what users see to what apps do across several fronts. On the UI side, researchers detect deceptive or misleading designs (dark patterns) directly from screenshots \cite{mansur2023aidui}, learn robust visual similarity metrics for Android screens \cite{li2024uihash}, and study permission–intent relationships via static analysis of layouts and code \cite{xi2019deepintent,yang2022describectx}, thereby improving how we read interfaces. On the behavioural side, privacy-compliance systems align claimed disclosures with actual flows, checking flow-to-policy consistency at scale \cite{tan2023ptpdroid}, or combine static flows with dynamic traffic to catch non-compliant transmissions \cite{fan2024giving}. In exploration and testing, LLM-driven agents now plan over UI state to drive realistic interactions and improve coverage \cite{liu2024make,feng2024prompting,wang2025llmdroid,yoon2024intent}, while runtime frameworks such as Guardian \cite{ran2024guardian} constrain the action space and offload navigation and domain rules to an external controller. Collectively, these advances strengthen UI understanding, compliance checking, and UI-guided exploration; however, they still fall short of providing a runtime, screen-level alignment between inferred user-visible intent and observed execution-time behaviour, including runtime semantics that can both steer exploration and treat misalignment itself as a first-class security signal.

In this paper, we introduce \uixpose{}, a GUI-driven analysis framework, which makes Intention-Behaviour Alignment (IBA) a primary runtime objective, aligning what the current screen implies the app intends to do with what the app actually does on the device and over the network during execution. \uixpose{} is explicitly designed to sit on top of advanced automated GUI testing frameworks, treating each screen both as a specification of user-visible intent and as a handle for driving systematic exploration. Concretely, \uixpose{} (a) infers screen-level intent with a vision-language model and a lightweight domain ontology (i.e., knowledge structure), (b) summarises execution-time behaviour, including decoded network payload semantics and resource usage, with provenance linked to the originating UI state, and (c) calculates a quantitative misalignment score that not only flags suspicious activity but also can be used to drive a closed-loop exploration policy toward behaviourally rich paths that conventional explorers overlook. This design addresses situations where permissions, resource usage, and endpoints appear harmless, but the payload content conflicts with the user-visible intent, or where malicious behaviours emerge only along specific interaction paths.

Intention–behaviour misalignment is a fundamental blind spot, meaning users (and analysts) reason from what the screen appears to do, yet many malicious behaviours manifest in background communications and resource use that the UI does not justify.

Prior static, dynamic, and hybrid detectors, as well as UI-only and privacy-compliance systems, either lack detailed runtime semantics or ignore what the current screen implies, rendering them unable to flag suspicious behaviour when endpoints and permissions appear benign. Still, payloads or resource patterns diverge from UI intent.

\uixpose{} turns this alignment into a first-class runtime signal. In three real-world apps, it surfaces covert exfiltration, hidden background activity, and instability patterns that simple coverage or metadata-based analyses miss, illustrating both the potential and limitations of this approach.

These observations yield four design goals derived directly from intention–behaviour alignment.

\textbf{G1 — Deep inspection.} Capture and condense runtime semantics (e.g., payload and resource) such that the behaviour summariser preserves information needed for alignment with UI intent, including decoded payload content and resource use, with provenance.

\textbf{G2 — Intent-aware Inference.}
Use the IBA signal to inform feedback as context for better interactions and screen inferences, aiming to increase expected misalignment and reduce false negatives due to under-exploration.

\textbf{G3 — Alignment-based detection.} Compute a quantitative misalignment score between inferred screen-level intent and execution-time behaviour; flag suspicious activity even when endpoints, permissions, or coarse indicators appear benign.

\textbf{G4 — Ensure high extensibility and applicability.} The framework should be modular and support plug-and-play integration of state-of-the-art components without modifying the IBA objective. It should also enable analysis of apps without requiring source code, while still utilising source code when available.

\uixpose{} realises G1 to G4 through four interdependent components centred on the IBA objective.  First, an instrumented runtime with an inline network path logs UI callbacks and resource usage in a sandbox environment, utilising a network proxy that captures and decodes payloads. It links flow provenance to the originating UI state (G1).

%an alignment-guided feedback engine that integrates advanced automated graphical user interface (GUI) testing frameworks enables automated exploration of app screens through generated user interactions.It incorporates the past feedback to infer richer behaviour or increase potential misalignment (G2). 
Second, a vision-language model-assisted alignment engine incorporates prior feedback to infer richer behaviour or to increase the potential for misalignment. It integrates advanced automated graphical user interface (GUI) testing frameworks, enabling automated exploration of app screens through generated user interactions (G2). Third, a screen-level intent module (vision–language with a lightweight ontology) and a behaviour summariser over execution traces project UI intent and observed behaviour into a shared action space, computing a quantitative misalignment score that flags suspicious activity (G3). 

Finally, \uixpose{} is modular, offering high extensibility without changing the IBA objective or event schema (G4). It employs a black-box analysis model, enabling application in real-world scenarios without requiring access to source code, while also supporting source-based analysis when available.

Section \ref{sec:motiv} presents the motivation behind designing \uixpose{}, followed by Section~\ref{sec:approach}, where we describe the detailed design of \uixpose{}. Section~\ref{sec:evaluation} presents our prototype and preliminary evaluations on real-world Android apps. In Section~\ref{sec:discussion}, we discuss the current limitations of the \uixpose{} module, which is built on existing techniques, and highlight the research challenges for its potential in mobile malware detection. Finally, we conclude and outline future research directions in Section~\ref{sec:conclusion}.

We next ground this high-level problem in a concrete motivating example, which exposes the misalignment between UI intent and runtime behaviour and directly motivates our design goals.

%% file: sections/motivation.tex
\section{Background and Motivation}
\label{sec:motiv}
Mobile malware has increasingly adopted techniques that frustrate traditional static and dynamic analyses. Code obfuscation, dynamic code loading, and reflective dispatch complicate static reasoning, while evasive runtime behaviours hide behind benign user interfaces and coarse system signals. While hybrid systems combine static and dynamic features, they still struggle to \emph{connect} user-visible intent to runtime semantics, and they often under-explore behaviourally rich paths.

We examined a popular Unity-based game (``Agent Shooter'') that exhibits invasive advertising and telemetry behaviours \cite{hiddnads}. Manual reverse engineering revealed heavy obfuscation and reflection (e.g., \texttt{DexClassLoader}, \texttt{invoke}, \texttt{getDeclaredClasses}, \texttt{setAccessible}), and dynamic loading of code modules-patterns known to degrade static precision~\cite{pendlebury2019tesseract,idrees2017pindroid,wu2023droidrl,zhu2023android,ksibi2024efficient}. Purely dynamic approaches were also challenged as Unity-driven UIs limited the effectiveness of generic GUI explorers~\cite{li2017droidbot,koroglu2018qbe}, reducing path coverage and the chance of triggering problematic states. 

\begin{figure}[!t]
    \centering
    \begin{minipage}{0.49\textwidth}
        \centering
        \includegraphics[keepaspectratio, scale=0.42, trim={0 0 0 0}, clip]{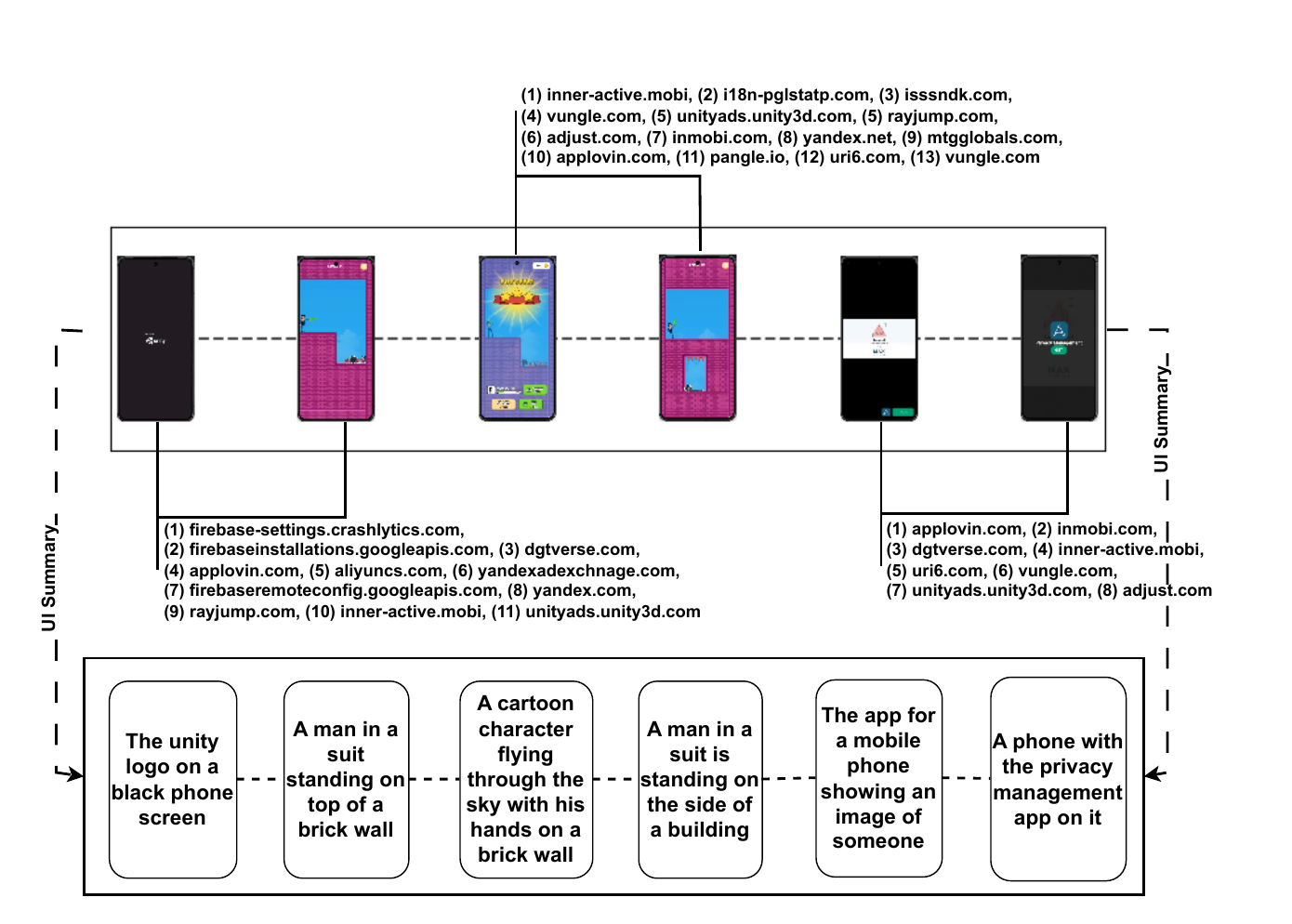}
        \subcaption{Underlying network communications with the corresponding UI screen and essential summary of each UI. Descriptive UI summaries for brevity and simplicity are removed.}
        \label{fig:res_pred_a}
    \end{minipage}
   \hfill
   \begin{minipage}{0.49\textwidth}
        \centering
        \includegraphics[keepaspectratio, scale=0.44, trim={0 0 0 0}, clip]{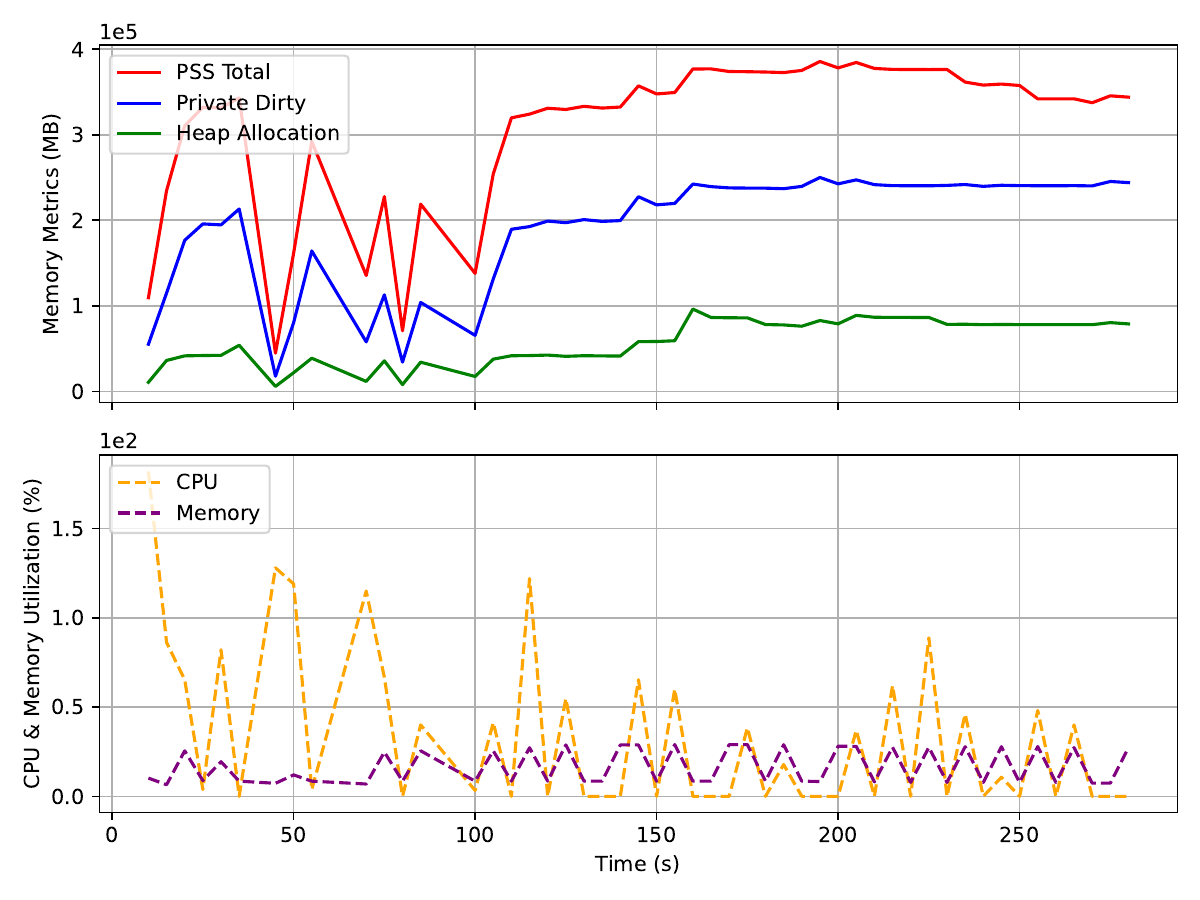}
        \subcaption{Resource utilisation from the perspective of memory usage (top) and utilisation (bottom).}
        \label{fig:res_pred_b}
    \end{minipage}
    \caption{Agent Shooter app's UI alignment analysis with the underlying network communications and resource utilisation.}
    \label{fig:res_pred}
\label{fig:agentshooter}
\end{figure}

% \paragraph{What users see vs.\ what the app does}
Screens in the game frequently display idle or menu-like content, yet the runtime exhibits bursts of background communication and resource use. Figure~\ref{fig:res_pred_a} (UI-to-communications view) illustrates screens whose UI suggests no network-heavy action, while the app communicates with a broad set of ad-tech/analytics endpoints. Figure~\ref{fig:res_pred_b} (resource profile) shows corresponding CPU/memory spikes early in interaction, settling into a sustained plateau despite a visually static UI. These observations are not just ``anomalies'' in isolation; they are instances of \emph{misalignment}, meaning the behaviour implied by the UI (intent) diverges from the behaviour observed at runtime (network payloads, heap/memory, CPU/IO).

% \paragraph{Why prior techniques miss this}
Static, permission-centric intent detectors reason about what the app \emph{could} do but not what it \emph{did} on this screen. Dynamic systems that rely on broad metadata (endpoints, IPs, aggregate CPU) cannot judge whether behaviour is \emph{appropriate} for the current UI, and are susceptible to evasion via IP churn, encryption, or benign-looking domains. Coverage-first GUI explorers can still miss states in which interesting behaviour manifests.

% \paragraph{Design goals that follow}
These limitations motivate four goals that shape \uixpose{}:
\emph{(G1) Deep inspection.} Capture multi-signal runtime semantics (decoded traffic where permissible, heap/memory, CPU/IO) and condense them into a common action space so alignment is possible.
\emph{(G2) Intent-aware exploration.} Use misalignment novelty and backend magnitude to increase expected misalignment and reveal richer behaviour under a fixed budget.
\emph{(G3) Alignment-based detection.} Compare a screen-level intent vector (from UI summarisation) with the fused behaviour vector; flag low-alignment/high-magnitude states, even when endpoints and permissions look benign.
\emph{(G4) Ensure high extensibility and applicability.} Operate on APKs without source; exploit source when available.

The case study demonstrates that merely collecting more runtime signals or triggering additional UI events is insufficient; what is required is a principled \emph{intention--behaviour alignment} check. \uixpose{} instantiates this by inferring screen intent with a vision--language pipeline, fusing multi-channel runtime evidence into a behaviour vector, and using alignment both to detect suspicious states. The result is a detector and explorer that reflect what the user sees, and how the app actually behaves, on each screen.

Taken together, these systems improve static reasoning, runtime monitoring, UI understanding, and policy compliance, but they still lack a runtime, screen-level alignment between what the UI currently promises and what the app actually does (including payload content and resource dynamics). None of them uses misalignment itself to steer exploration toward behaviourally rich paths. This gap motivates the design of \uixpose{} as an IBA-first framework.

The following section instantiates these goals in the \uixpose{} framework, showing how each architectural component is explicitly designed to deliver deep inspection (G1), intent-aware exploration (G2), alignment-based detection (G3), and modular applicability (G4).

%% file: sections/approach.tex
\section{Framework Design}\label{sec:approach}

\begin{figure}[!t]
  \centering
  \includegraphics[width=1.0\linewidth, keepaspectratio]{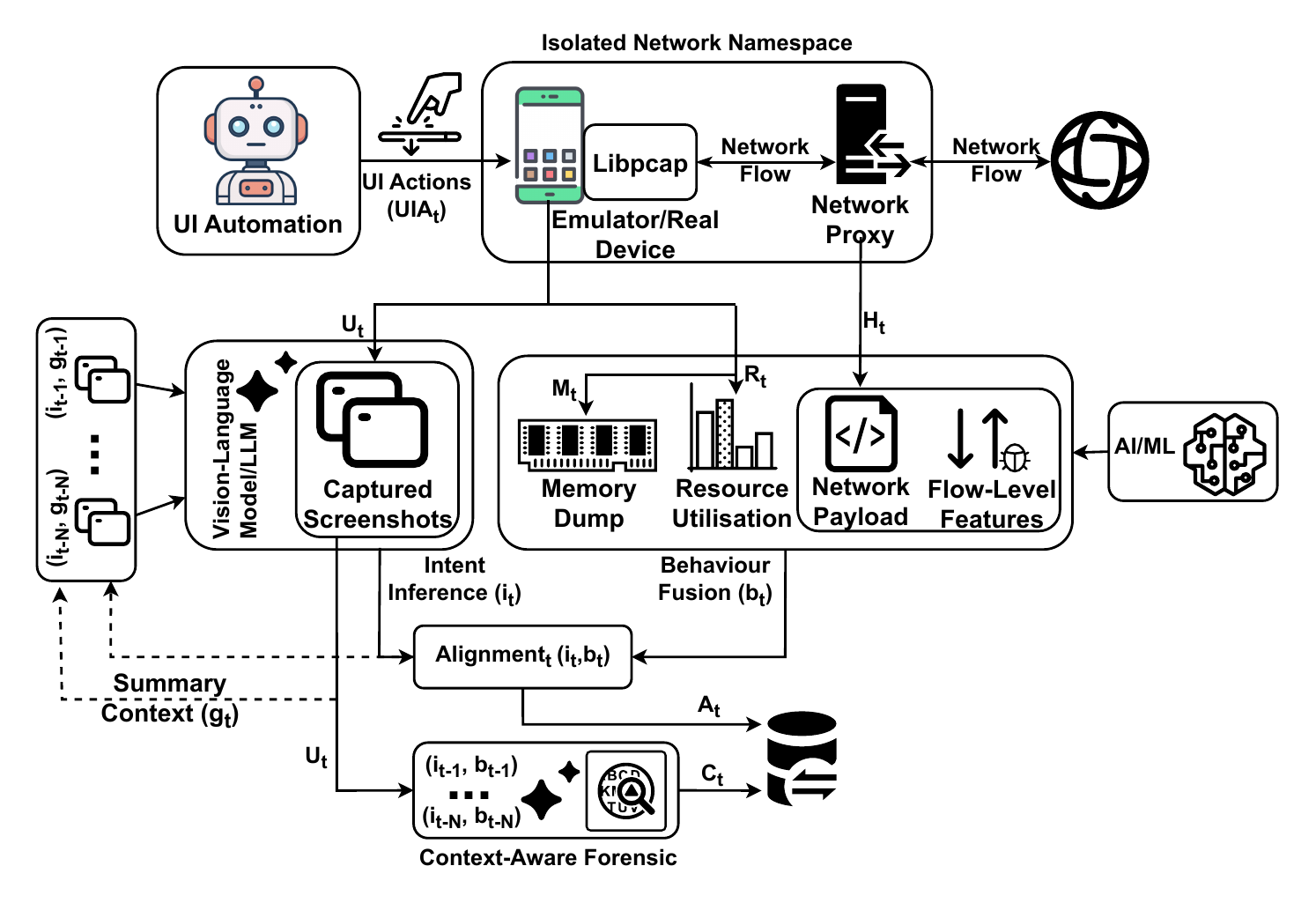}
  \caption{\uixpose{} overall design, and its key interdependent components. At time $t$, each UI action ${UIA}_t$ may incur network, memory, and resource utilisation, which builds $\mathcal{A}_t$ and its associated behaviour fusion $b_t$. The LLM assists with two aspects, considering the past $N$ inferences: (1) inferring an intent vector $i_t$ for alignment assessment $A_t$ and (2) outputting confidence $C_t$ based on screenshot context analysis, $A_t$, $i_t$, and $b_t$.}  \label{fig:uixpose}
\end{figure}

Building on the limitations and goals identified in Section \ref{sec:motiv}, we now describe the \uixpose{} framework, which operationalises intention–behaviour alignment at runtime.

\uixpose{} is a pluggable, modular system designed to automate the detection of malicious behaviour. Its overall architecture is shown in Figure~\ref{fig:uixpose}. \uixpose{} instruments an isolated device or emulator to collect time-synchronised screenshots, decoded traffic, heap snapshots, and resource utilisation traces assisted by UI automation tools that send a sequence of UI actions ($UIA_t$). From each screenshot $u_t$, a vision-language model (VLM)/LLM pipeline produces a calibrated intent vector $i_t$ over an action ontology $\mathcal{A}$, converted into per-action evidence vectors at time $t$ as $\{\mathbf{s}^{(M)}_t, \mathbf{s}^{(R)}_t, \mathbf{s}^{(H)}_t\}$, representing, the network utilisation, memory pressure, and resource intensity, assisted with artificial intelligence or machine learning techniques. \uixpose{} fuses evidence into a behaviour vector ($b_t$), and computes alignment score $A_t = sim(i_t, b_t)$ with a scalar fallback when only one shared dimension exists. The pipeline does not require the app's source code and is modular in practice, while maintaining alignment with the objective.  

\subsection{UI Automation and Intent Summarisation}
\label{sec:ui_auto_sum}

\uixpose{} is agnostic to the choice of UI automation explorer, but it enables the surfacing of diverse screens.
Therefore, \uixpose{} can benefit from state-aware explorers with programmatic hooks, such as DroidBot~\cite{li2017droidbot}, a reinforcement-learning (RL) strategy~\cite{pan2020reinforcement}, and Fastbot~\cite{fastbot2}, as it can be queried for the current UI state and driven by an external policy. In contrast, the Monkey tool~\cite{monkey_ui} is
a purely random event generator and cannot perform purposeful inputs (e.g., input entry); consequently, \uixpose{} retains Monkey as a baseline tool.

To compare UI intent with execution-time behaviour, \uixpose{} maps the current screen $u_t$ into an intent vector $\mathbf{i}_t \in [0,1]^{|\mathcal{A}|}$ over an action-primitive set $\mathcal{A}$. \uixpose{} may use a vision--language model (VLM), e.g., BLIP~\cite{li2022blip}, to caption $u_t$ and extract UI objects or use LLMs to enhance UI summarisation further, as LLMs can analyse UI components, layouts, and features to produce concise, natural-language descriptions. 

LLM prompting for intent inference starts with giving a screenshot, the LLM is prompted to enumerate salient UI elements and on-screen hints, considering a concise summary of the previous $N$ LLM-produced intent vectors ($i_t$) and corresponding summary goals ($g_t$) as the \emph{context} [($i_{t-1}$, $g_{t-1}$), ...,($i_{t-N}$, $g_{t-N}$)], and then return a calibrated intent vector $\mathbf{i}_t \in [0,1]^{|\mathcal{A}|}$ over an extendable action primitives $\mathcal{A}$. The prompt constrains the model to a compact schema with bounded probabilities and a short rationale, and instructs it to rely only on visible screen evidence and the prior knowledge from previous steps. 

These models aim to map UI phrases to $\mathcal{A}$, in which the result is $\mathbf{i}_t$, where each
dimension encodes the probability that the current screen intends the corresponding action.

The explorer exposes a set of feasible interactions. After each action, \uixpose{} computes alignment or misalignment (G2). Let ${t:t+k}$ be the execution trace window following $u_t$. A behaviour summariser maps the trace into a behaviour vector $\mathbf{b}_t \in [0,1]^{|\mathcal{A}|}$ by combining three channels, which are (a) Network Flow \& Payload Profiler (NFPP), referred to as $H_t$ channel for \texttt{network utilisation}, which consumes per-flow statistical features and decoded payload cues to produce probabilities over network-driven actions; (b) Memory Pressure Profiler (MPP), known as $M_t$ channel for \texttt{memory pressure}, which consumes \emph{meminfo}-derived levels and $\Delta$-time features (e.g., $\Delta$HeapAlloc/$\Delta t$), to produce probabilities for memory/Inter Process Communication (IPC) pressure, and (c) Resource Utilisation Profiler (RUP) as $R_t$ channel for \texttt{resource intensity} consumes scheduler-level process metrics (e.g., from \texttt{top}: CPU\%, MEM\%) and their $\Delta/\Delta t$) to produce probability of resource-intensity. Therefore, the action space is defined as $\mathcal{A}=\{H_t, M_t, R_t\}$.

%detector
Each channel produces a per-action evidence score $\mathbf{h}_t$, $\mathbf{m}_t$,
$\mathbf{r}_t \in [0,1]^{|\mathcal{A}|}$. \uixpose{} constructs the aggregated behaviour vector $\mathbf{b}_t$ as
\begin{equation}
\begin{split}
\mathbf{b}_t \;=\; \frac{\sum_{k\in\{H,M,R\}} \alpha_k\,\mathbf{s}^{(k)}_t}
       {\sum_{k\in\{H,M,R\}} \alpha_k \cdot m_k}, 
\\
\quad \mathbf{s}^{(H)}_t=\mathbf{h}_t,\; \mathbf{s}^{(M)}_t=\mathbf{m}_t,\;
\mathbf{s}^{(R)}_t=\mathbf{r}_t.
\label{eq:behfusion}
\end{split}
\end{equation}
where $\alpha_k$ are fusion weights and missing channels are masked via the indicator $m_k \in \{0,1\}$. 

\uixpose{} defines IBA at time $t$ using a geometry-based approach to handle vector instability at low magnitudes. 
Instead of cosine similarity, which is undefined for zero vectors, 
\uixpose{} calculates $A_t$ using a Radial Basis Function (RBF) kernel based on the squared Euclidean distance between the intent ($\mathbf{i}_t = [i_{net}, i_{mem}, i_{res}] \in [0,1]^3$) and behaviour ($\mathbf{b}_t$) vectors:

\begin{equation}
A_t \;=\; \exp\left(-\gamma \, \|\mathbf{i}_t - \mathbf{b}_t\|_2^2\right) \in (0, 1],
\quad M_t \;=\; 1 - A_t.
\label{eq:alignment}
\end{equation}
%(e.g., $\gamma=5.0$)
where $\gamma$ is a sensitivity hyperparameter. This provides a stable alignment score when both $\mathbf{b}_t$ and $\mathbf{i}_t$ are small (i.e., system noise), the distance approaches zero and $A_t \to 1$ (safe), effectively suppressing false positives during idle states, and suspicious activities can be flagged when $M_t \ge \theta$.
%A detector flags
For user-facing summaries,\uixpose{} reports the backend magnitude $B_t$ using the Euclidean ($L_2$) norm: 
% rather than an average:
\begin{equation}
B_t \;=\; \|\mathbf{b}_t\|_2 \ge 0,
\label{eq:backendmag}
\end{equation}
which captures the total energy of the observed behaviour without dilution. This allows single-channel anomalies (e.g., network-only beaconing) to be recorded with high magnitude. 

A 2D view $(A_t, B_t)$ aids triage with the following interpretation:
\begin{itemize}
    \item \textbf{High $B_t$, High $A_t$ (Safe/Authorised):} The application is under heavy load, but the activity matches the predicted intent. %(e.g., video streaming).
    \item \textbf{High $B_t$, Low $A_t$ (Overt Anomaly):} High-energy deviation. The application performs intensive, unpredictable work (e.g., exfiltration).
    \item \textbf{Low $B_t$, High $A_t$ (Idle/Safe):} Background noise where behaviour and intent are both negligible.
    \item \textbf{Low $B_t$, Low $A_t$ (Uncertain/Stealth):} Minor unpredicted fluctuations or low-volume stealth activity.
\end{itemize}

\subsubsection{Context-Aware Forensic Verification.}
While vector alignment detects directional mismatches, it cannot detect \textit{semantic spoofing}, e.g., an app masking high network activity behind a generic ``Loading" intent. To resolve this, \uixpose{} consults with LLMs as a \emph{forensic judge} $\Phi$ for the final step. The judge receives the current screenshot $\mathcal{I}_t$, the behaviour vector $\mathbf{b}_t$, and the intent prediction $\mathbf{i}_t$ and its $N$ previous intent vectors, i.e., [($i_{t-1}$, $b_{t-1}$), ..., ($i_{t-N}$, $b_{t-N}$)] to compute a verification confidence $C_t$:

\begin{equation}
C_t = \Phi(\mathcal{I}_t, \mathbf{b}_t, \mathbf{i}_t, [(i_{t-1}, b_{t-1}), ...,(i_{t-N}, b_{t-N})]) \in [0,1]
\end{equation}

The judge applies a three-layer logic to validate the dominant active channel in $\mathbf{b}_t$:
\begin{itemize}
    \item \textbf{Layer 1 - Visual Evidence:} Checks for explicit UI indicators justifying the load (e.g., spinners for Network, keyboards for CPU).
    \item \textbf{Layer 2 - Implicit Context:} Validates non-visual but authorised operations, considering the previous behaviour and intention vectors that led to this screen.
    \item \textbf{Layer 3 - Anomaly Trap:} Detects logical contradictions and outputs the confidence level, given the context.
\end{itemize}
This produces a final classification that flags events with High $A_t$ that are mathematically aligned but Low $C_t$ as visually unverified.

\subsection{UI-Conditioned Intent Inference from Visual Evidence (\texttt{$i_t$})}

%an action–space intent vector $i_t = [i_{net}, i_{mem}, i_{res}] \in [0,1]^3$
\uixpose{} infers the action–space intent vector $i_t$ from a screenshot in two steps. First, Algorithm~\ref{alg:ui_to_evidence} involves a vision model that emits an evidence JSON containing detected UI components (type, bounding box, confidence) and state cues, such as spinner/progress indicators or error/empty banners. This evidence incorporates the previous $N$ tuples of [($i_{t-1}$, $g_{t-1}$),...,($i_{t-N}$, $g_{t-N}$)] as the context for better analysis. This is because UIs are not inherently independent; in fact, a sequence of past actions leads to the new UI state. 

We treat per-component weights $w_t$, and effects $E[t]$ as heuristic priors reflecting domain knowledge about the typical backend impact of visible UI elements to be used in Alg.~\ref{alg:evidence_to_intent}.  \uixpose{} employs Alg.~\ref{alg:evidence_to_intent} and applies a \textit{deterministic} map, per-component impacts (product weights $w_t$ and $E_t$) are accumulated, softly capped, squashed via a sigmoid, and adjusted by visible state modifiers, yielding $i_t$. This produces an auditable, reproducible prior for downstream alignment with backend evidence.

Alg.~\ref{alg:evidence_to_intent} turns a single Android UI screenshot into an auditable JSON “evidence pack” that a downstream model can trust. First, a vision-language model detects only components from a fixed ontology with respect to the RICO-defined UI classes \cite{deka2017rico} (e.g., `Toolbar') and outputs, for each detection, its class, bounding box, any visible text, and a calibrated confidence score. A short screen summary is formed strictly from what is on the screen. Finally, everything is emitted as strict JSON with validation (classes, well-formed boxes, and correct JSON types). Low-certainty items are either omitted or explicitly marked with low confidence, allowing downstream logic to handle uncertainty.

Within this framework, ``state indicators" are compact flags or measurements that describe the current condition of the screen, derived solely from visual evidence. They summarise what is happening using components; for example, a rotating ring is displayed on screen to represent loading. Each indicator should carry a confidence and minimal evidence (e.g., the bounding box and the quoted text) so that every claim is traceable back to pixels. In short, components explain ``what'' is on the screen; state indicators present what it means right now for interaction and app behaviour.

\begin{algorithm}[!t]
\small
\DontPrintSemicolon
\SetAlgoLined
\SetKwInOut{KwIn}{Input}\SetKwInOut{KwOut}{Output}
\SetKw{Return}{return}
\SetKwFunction{Clamp}{Clamp}
\SetKwFunction{Sigmoid}{Sigmoid}

\caption{UI $\rightarrow$ EVIDENCE }
\label{alg:ui_to_evidence}

\KwIn{Android UI screenshot $\mathcal{I}$ and [($i_{t-1}$, $g_{t-1}$), ...,($i_{t-N}$, $g_{t-N}$)]}
\KwOut{$\textsf{EVIDENCE}$ JSON with fields:
$\textsf{components}$, $\textsf{state\_indicators}$, $\textsf{screen\_summary}$}
%, \texttt{Checkbox}
\BlankLine
\textbf{State Indicators} \& \textbf{Component Ontology} $\mathcal{T}$: \{\texttt{BackgroundImage}, \texttt{Bottom\_Navigation}, \texttt{Card}, \texttt{CheckBox(box)}, \texttt{CheckedTextView}, \texttt{Drawer}, \texttt{EditText}, \texttt{Icon}, \texttt{Image}, \texttt{Map}, \texttt{Modal}, \texttt{Multi\_Tab}, \texttt{PageIndicator}, \texttt{Remember}, \texttt{Spinner}, \texttt{Switch}, \texttt{Text}, \texttt{TextButton}, \texttt{Toolbar}, \texttt{UpperTaskBar}\}\;

\BlankLine
\textbf{1. Visual Detection (VLM call, evidence-only).}\;
\Indp
  a) Detect instances of types in $\mathcal{T}$ on $\mathcal{I}$; for each detection $c$, output:
  \(\textsf{type}\in\mathcal{T}\), \(\textsf{bbox}=[x_{\min},y_{\min},x_{\max},y_{\max}]\),
  \(\textsf{visible\_text}\), and \(\textsf{confidence}\in[0,1]\).\;
  b) Extract \textbf{state indicators} only if visually supported (booleans and exact on-screen strings).\;
\Indm

\textbf{2. Screen Summary (auditability).}\;
\Indp
  Produce a short description, primary user goal (must cite visible labels), and 2--5 evidence bullets
  (component type + bbox + text).\;
\Indm

\textbf{3. JSON Emission (strict).}\;
\Indp
  Assemble $\textsf{EVIDENCE}=\{\textsf{components}:[\cdot],\ \textsf{state\_indicators}:\{\cdot\},\
  \textsf{screen\_summary}:\{\cdot\}\}$; omit uncertain items or include with low confidence.\;
  Validate: allowed types, integer bboxes with $x_{\min}{<}x_{\max}$, $y_{\min}{<}y_{\max}$;
  booleans as JSON; numbers as JSON.\;
\Indm

\Return $\textsf{EVIDENCE}$\;
\end{algorithm}

% Deterministic Scoring: 
\begin{algorithm}[!t]
\small
\DontPrintSemicolon
\SetAlgoLined
\caption{EVIDENCE $\rightarrow$ $\mathbf{i}_t$}
\label{alg:evidence_to_intent}
\KwIn{$\textsf{EVIDENCE}$; component params $w_t$, $E[t]$; caps $\kappa$; scales $\tau$;\\
\hspace{1.8em} state params $\alpha_{a,k}\!\ge\!0$, $\beta_{a,k}\!\in\![0,1]$ for $a\!\in\!\{\mathrm{net,mem,res}\}$ and indicators $k\!\in\!\mathcal{K}$;\\
\hspace{1.8em} keyword weights $\theta_w\!\ge\!0$ for text indicators.}
\KwOut{$\mathbf{i}_t=[i_{\text{net}},i_{\text{mem}},i_{\text{res}}]\in[0,1]^3$}

\textbf{Init:} $s_{\text{net}},s_{\text{mem}},s_{\text{res}}\leftarrow 0$.\;

\textbf{Per-component accumulation:} \ForEach{$c\in \textsf{components}$}{
  $t\leftarrow c.\textsf{type}$; $p\leftarrow \mathrm{clip}(c.\textsf{confidence},0,1)$\;
  \If{$t\notin w_t$ \textbf{ or } $t\notin E$}{\textbf{continue}}
  $[e_{\text{net}},e_{\text{mem}},e_{\text{res}}]\leftarrow E[t]$\;
  $s_{\text{net}}{+}{=}w_t[t]\cdot p\cdot e_{\text{net}}$;\;
  $s_{\text{mem}}{+}{=}w_t[t]\cdot p\cdot e_{\text{mem}}$;\;
  $s_{\text{res}}{+}{=}w_t[t]\cdot p\cdot e_{\text{res}}$\;
}

\textbf{Soft caps \& squash:}
$s_a\leftarrow \min(s_a,\kappa_a)$ for $a\in\{\text{net},\text{mem},\text{res}\}$;\;
$i_a\leftarrow \sigma(s_a/\tau_a)$ with $\sigma(x){=}1/(1{+}e^{-x})$.\;

\textbf{Indicator severities:} $S\leftarrow \textsf{state\_indicators}$;\; 
\ForEach{$k\in\mathcal{K}$}{
  compute $m_k\in[0,1]$ from $S$ (booleans $\rightarrow \{0,1\}$; 
  ratio $\rightarrow 1{-}\texttt{progress\_determinate\_ratio}$; 
  text $\rightarrow \mathrm{clip}(\sum_{w}\theta_w\,\mathbf{1}[w\in \text{text}],0,1)$; 
  media $\rightarrow \{1,0.5,0\}$).
}

\textbf{State composition:}
$\Delta^{\text{add}}_a \leftarrow \sum_{k}\alpha_{a,k}\, m_k$;\quad
$\Gamma^{\text{mul}}_a \leftarrow \prod_{k}\left(1-\beta_{a,k}\, m_k\right)$.\;

\textbf{Apply \& return:}
$i_a \leftarrow \mathrm{clip}\!\left(i_a\cdot \Gamma^{\text{mul}}_a + \Delta^{\text{add}}_a,\,0,1\right)$;\;
\Return $[i_{\text{net}},i_{\text{mem}},i_{\text{res}}]$.\;
\end{algorithm}

\subsection{Isolated Runtime \& Time-Synchronised Telemetry (G4)}
\label{sec:runtime}

\uixpose{} executes each target app on either a dedicated Android emulator or a tethered smartphone; in both cases, the device endpoint is placed inside a separate Linux network namespace. The namespace isolates traffic, allowing only packets to and from the device to be visible to the framework, and removes host noise, which improves the attribution of measurements to the app under test.

\paragraph{Network capture}
On the emulator built from source code, \texttt{libpcap} is integrated and enabled on the virtual interface inside the namespace. On physical devices, we route traffic through a user-space VPN (OpenVPN) that terminates within the namespace, resulting in a device-specific flow, and interprets network flows with a standalone \texttt{libpcap} script. Capture occurs at the IP layer, and when possible, a proxy root certificate enables TLS interception for HTTP(s) decoding; otherwise, informative metadata (e.g., service name indication (SNI), or sizes) is still retained with the option to keep all network flows in the form of \texttt{pcaps} for reproducibility.

%with OCR text
\paragraph{Resource, heap, and UI capture}
A lightweight orchestrator monotonically timestamps and aligns periodic resource utilisation telemetry (from \texttt{top}/\texttt{dumpsys}); heap snapshots (e.g., \texttt{dumpsys meminfo}), and screenshots\footnote{Memory objects information is discussed in Android official developer guide \cite{dumpsys}.}. Sampling uses a uniform interval $\Delta t$ and produces a per-step bundle $\langle \text{screenshot}, \text{pcap window}, \text{resource slice}, \text{heap stats}\rangle$ for alignment. This setup provides low-impact monitoring and ensures all signals are attributable and comparable at step~$t$.

\subsection{Multi-Signal Evidence Extractors (G1)}
\label{sec:evidence}

Each telemetry channel emits a vector $\mathbf{s}^{(k)}_t \in  [0,1]^{|\mathcal{A}|}$ that estimates how strongly the channel supports each action primitive at step $t$.
% that is $\mathcal{A}=\{\texttt{net-utilisation},\texttt{mem-pressure},\texttt{res-intensity}\}$.

\vspace{0.35em}
\noindent\textbf{NFPP-H: Network semantics via feature and flow fusion.}
At step $t$, \uixpose{} aligns a sliding window $[t{-}\Delta,t]$ and parses two complementary sources.

\emph{Payload path HTTP(s).} If TLS interception is available, \uixpose{} decodes bodies; otherwise, it operates header/flow-only. 
For each request/response, \uixpose{} computes (a) metadata-robust signals, e.g., method rarity; header presence atypicality; status-class deviation; total message size; (b) tracking indicators, e.g., third-party/cross-domain relative to the app’s registrable origin; tracker/domain matches via curated lists and reputation feeds; and (c) behavioural correlations. \uixpose{} avoids naive payload entropy when encryption is used; under interception, it includes body entropy as an auxiliary feature. Network flows receive a reputation score from a classifier fed by libpcap-captured features. Each message gets a capped, weighted score $s \in [0,1]$. 
If the window is empty, the H-channel is masked ($m_H{=}0$), else $m_H{=}1$.

\emph{Flow–ML path.} The libpcap-based CNN pipeline emits \emph{per-interval snapshots} with cumulative bytes per flow (\texttt{srcIP-dstIP-srcPort-dstPort}). For each row, cumulative totals are compared to obtain per-step bytes and mark the flow as \emph{malicious} iff a symmetric match exists ($\hat p_f{=}1$). We aggregate malicious volume and normalise it at the session level below, so brief beacons do not dominate while sustained risky throughput is reflected.

From payload/matching, \uixpose{} forms per-step rates as bytes $B_t$, requests $R_t$, failures $F_t$, endpoint-reputation $E_t$, tracker-density $T_t$, and an encode/decode proxy $H_t$ (entropy $\times$ size). \uixpose{} also computes a cross-domain ratio $X_t$, the large-payload share $P_t$, and an in-flight backlog proxy $L_t$ that includes unmatched/expired requests. For flows, \uixpose{} obtains a malicious-bytes rate $M^{(\mathrm{bytes})}_t$ by summing per-flow deltas. All series are smoothed via Exponentially Weighted Moving Averages (EWMAs) with a smoothing factor $\alpha$ and normalised by a session-local rolling maximum over an adjustable steps ($W$). The per-step H-evidence vector $\mathbf{m}^{(H)}_t$ projects onto $\mathcal{A}$ as

\begin{equation}
\begin{split}
m^{(H)}_{\mathrm{net}}(t)= \min\!\Big\{
  w_b\,\widetilde{B}_t + w_r\,\widetilde{R}_t + w_f\,\widetilde{F}_t \\
+ w_x\,\widetilde{X}_t +  
w_e\,\widetilde{E}_t + w_t\,\widetilde{T}_t 
+ w_m\,\widehat{M}_t,\; 1 \Big\}, \\[4pt]
m^{(H)}_{\mathrm{mem}}(t)
= \min\!\Big\{ \kappa_L\,\widetilde{L}_t + \kappa_P\,\widetilde{P}_t,\; MAX_{m^{(H)}_{\mathrm{mem}}(t)} \Big\}, \\[4pt]
m^{(H)}_{\mathrm{res}}(t)
= \min\!\Big\{ \kappa_H\,\widetilde{H}_t,\; MAX_{m^{(H)}_{\mathrm{res}}(t)}\Big\}.
\end{split}
\end{equation}

where $(\kappa_L,\kappa_P,\kappa_H)$ are the spillover multipliers for the non-network axes, ensuring non-network axes signals nudge decisions (e.g., retries or backlog) without overwhelming network throughput/rate evidence. If the H-channel lacks sufficient samples, the output is masked as  $\mathbf{m}^{(H)}_t=\mathbf{0}$.
The flow term $\widehat{M}_t$ adds coverage when payloads are opaque/absent and is bounded ($w_m$) to avoid double-counting with throughput. 

\vspace{0.35em}
\noindent\textbf{MPP-M: Heap/memory semantics.}
At each step $t$, \uixpose{} parses \texttt{dumpsys meminfo} and forms a counter vector $\mathbf{s}^{(M)}_t$.
Rates are computed by finite differences over the sampling period $\Delta t$, then EWMA smoothing and session-local rolling-max normalisation map features to $[0,1]$. Let $\dot{PSS}_t$, $\dot{H}_t$, and $\texttt{heap\_pressure}_t$ denote by $\widetilde{(\cdot)}$ the EWMA and by $\mathrm{norm}(\cdot)\!\in\![0,1]$ the rolling-max normalisation.
\begin{equation}
\begin{split}
\dot{PSS}_t=\frac{\texttt{PssTotal}_t-\texttt{PssTotal}_{t-1}}{\Delta t},\\
\dot{H}_t=\frac{\texttt{HeapAlloc}_t-\texttt{HeapAlloc}_{t-1}}{\Delta t},\\
\texttt{heap\_pressure}_t=\frac{\texttt{HeapAlloc}_t}{\texttt{HeapSize}_t+\varepsilon}.
\end{split}
\end{equation}

\begin{equation}
\begin{split}
\label{eq:m-mem}
m^{(M)}_{\text{mem}}(t)
=
\Big[
\alpha_1\,\mathrm{norm}\!\big(\widetilde{\dot{PSS}}_t\big)+\\
\alpha_2\,\mathrm{norm}\!\big(\widetilde{\dot{H}}_t\big)+
\alpha_3\,\mathrm{norm}\!\big(\texttt{heap\_pressure}_t\big)+\\
\alpha_4\,\mathrm{norm}\!\big(\widetilde{\texttt{SwapPssDirty}}_t\big)
\Big]\cdot \mathbf{1}\{\overline{u}_t\ge \tau_{\text{mem}}\},\\
\overline{u}_t=\frac{1}{K}\sum_{k=0}^{K-1}\mathrm{EWMA}\!\left(\text{raw }m^{(M)}_{\text{mem}}(t-k)\right).
\nonumber
\end{split}
\end{equation}

\uixpose{} treats network and computes projections conservatively as small handshake proxies that may raise \emph{$H$} slightly via binder/webview rates, while computes reflects UI inflation churn. Define
\begin{equation}
\begin{split}
r^{\text{binder}}_t=
% \frac{\Delta(\texttt{LocalBinders}+\texttt{ProxyBinders}+\texttt{ParcelCount})}{\Delta t},\\
\frac{\Delta(\texttt{LocalBinders}+\texttt{ProxyBinders}}{\Delta t}+\\ \frac{\Delta(\texttt{ParcelCount})}{\Delta t},
r^{\text{wv}}_t=\frac{\Delta \texttt{WebViews}}{\Delta t},\\ \quad
r^{\text{ui}}_t=\frac{\Delta(\texttt{Views}+\texttt{ViewRootImpl})}{\Delta t}.
\end{split}
\end{equation}

%into $\mathcal{A}$ 
Then the per-action vector contributed by the \emph{heap} is

\begin{equation}
\begin{split}
\mathbf{s}^{(M)}_t=\big[
\underbrace{\min\{1,\ \beta_{\text{ipc}}\mathrm{norm}(\widetilde{r}^{\text{binder}}_t)+\beta_{\text{wv}}\mathrm{norm}(\widetilde{r}^{\text{wv}}_t)\}}_{\text{H (handshake proxies)}},\; \\
\underbrace{m^{(M)}_{\text{mem}}(t)}_{\text{M}},\; 
\underbrace{\gamma_{\text{ui}}\mathrm{norm}(\widetilde{r}^{\text{ui}}_t)}_{\text{R (UI churn)}}\big],
\end{split}
\end{equation}

%All scalars are clamped to $[0,1]$. 
If \texttt{meminfo} is unavailable, the step is skipped (no mask emitted).

\vspace{0.35em}
\noindent\textbf{RUP-R: Resource signal/evidence.}

\label{sec:resource-evidence}
For each step $t$ , \uixpose{} aggregates process telemetry from \texttt{top} (and \texttt{/proc}) including \texttt{CPU\%}, \texttt{MEM\%},  Resident (\texttt{RES}), Virtual (\texttt{VIRT}), and Share (\texttt{SHR}). We explicitly normalise CPU ($\text{cpu-cap-norm}_t$) by device capacity using the number of logical cores $cr$.

% We retain a short window (default $w{=}5$) aligned to the IBA step and use ($w$)
\uixpose{} retains a window aligned with the IBA step and uses session-local rolling maxima for normalising levels and rates; volatility is captured via a rolling $z$-score. The compute scalar is dominated by core-aware CPU magnitude with a modest volatility term:
\begin{equation}
\begin{split}
m^{(R)}_{\text{res}}(t)=\mathrm{clip}\Big(\gamma_1\,\mathrm{norm}(\text{cpu-cap-norm}_t)\\
+\gamma_2\,\mathrm{zabs}(\texttt{CPU\%})_t,\ 0,1\Big).
\end{split}
\end{equation}

\uixpose{} maps resident set size (RSS) dynamics to a memory-pressure corroborator using rates-first with a sustained-elevation gate:
\begin{equation}
\begin{split}
\dot{m}_t=\frac{\texttt{MEM\%}_t-\texttt{MEM\%}_{t-1}}{\Delta t},
\dot{r}_t=\frac{\texttt{RES}_t-\texttt{RES}_{t-1}}{\Delta t},
\\
\dot{s}^{-}_t=\max\Big\{0,\ -\frac{\texttt{SHR}_t-\texttt{SHR}_{t-1}}{\Delta t}\Big\},
\\
m^{(R)}_{\text{mem}}(t)
=\Big[\alpha_1\,\mathrm{norm}(\mathrm{EWMA}(\dot{m}_t))
+\alpha_2\,\mathrm{norm}(\mathrm{EWMA}(\dot{r}_t))
\\
+\alpha_3\,\mathrm{norm}(\mathrm{EWMA}(\texttt{MEM\%}_t))
+\alpha_4\,\mathrm{norm}(\mathrm{EWMA}(\dot{s}^{-}_t))\Big] 
\\
\cdot \mathbf{1}\{\overline{v}_t\ge \tau^{(R)}_{\text{mem}}\},
\end{split}
\end{equation}
% with $(\alpha_1,\alpha_2,\alpha_3,\alpha_4)=(0.35,0.25,0.30,0.10)$, gate window $K{=}3$, and $\tau^{(R)}_{\text{mem}}{=}0.12$.

\texttt{top} exposes no byte/packet counters; to preserve orthogonality and avoid speculative proxies, \uixpose{} sets $m^{(R)}_{\text{net}}(t){=}0$. Thus, the resource signals contributes to $\mathcal{A}$ as

\begin{equation}
\mathbf{s}^{(R)}_t=\big[\,\underbrace{0}_{\text{H}},\ \underbrace{m^{(R)}_{\text{mem}}(t)}_{\text{M}},\ \underbrace{m^{(R)}_{\text{res}}(t)}_{\text{R}}\,\big].
\end{equation}

CPU-derived signals appear only in the \emph{R} projections; memory pressure is sourced from \texttt{meminfo} and, separately, from scheduler-side resource signals as a corroborator. \uixpose{} does not infer network load from scheduler metrics; the network/\texttt{meminfo} pipeline provides actual $H$, ensuring orthogonality and clean fusion.

Having defined the architecture and alignment mechanisms, we now instantiate \uixpose{} in a prototype implementation and conduct a preliminary evaluation to assess its feasibility and behaviour on real-world apps.

%% file: sections/evaluation.tex
\section{Preliminary Evaluation}\label{sec:evaluation}

Building on the framework in Section \ref{sec:approach}, \uixpose{}’s core analysis engine is implemented as a modular Python system that integrates an LLM for screen interpretation, manages UI interactions, and collects execution traces for alignment analysis. We developed \uixpose-frontend, a lightweight Django-based web interface for interacting with the engine and visualising results, and conducted a preliminary set of experiments to evaluate the performance of \uixpose{}\footnote{All reproducibility and validity artifacts are provided in the project repository.}. 

Section \ref{sec:ui_impact_hyperparameter} initialises the UI-impact prior and the IBA hyperparameters that parameterise the intent vectors and alignment score. Sections \ref{sec:experiment_res} and \ref{sec:casestudies_res} then evaluate \uixpose{} under this configuration on three real-world apps.

\subsection{UI-Impact Prior \& $\mathcal{A}$ Hyperparameters Initialisation}
\label{sec:ui_impact_hyperparameter}
% \hl{no dataset was available to provide the \texttt{top/ps ax} process level info, the gap, mention this}
To minimise single-expert bias and ensure the prior is auditable, we employ structured expert elicitation across five independent LLMs (the most advanced models) following a fixed schema and value bounds. We then apply robust aggregation for each parameter and class using a Huber M-estimator as the primary estimate, with median and 10\%/20\% trimmed means serving as sensitivity checks. Dispersion is reported via the interquartile range (IQR), and any parameters showing high disagreement are highlighted, yielding an interpretable, consensus prior. We present both the raw and final UI-impact prior, including the impact weight $w_t$ and effect values $E_t = (E_{net}, E_{mem}, E_{res})$, in Table \ref{tab:model_comparison}.%   
\begin{table*}[ht]
\centering
\small
\caption{Expert-elicited UI-impact prior: channel effect vectors and weights per surrogate; $w_t$ and $E[t]=[E_{\text{net}},E_{\text{mem}},E_{\text{res}}]$ across LLMs.}
\label{tab:model_comparison}
\footnotesize
\setlength{\tabcolsep}{3pt}
\renewcommand{\arraystretch}{0.95}
\resizebox{\textwidth}{!}{%
\begin{tabular}{@{}lccccc|c@{}}
\toprule
\multicolumn{6}{c|}{\textbf{LLMs}} & \textbf{UI-impact prior} \\ 
\cmidrule(lr){1-6} \cmidrule(lr){7-7}
\thead{Component} & \thead{ChatGPT} & \thead{Claude} & \thead{DeepSeek} & \thead{Gemini} & \thead{Llama} & \thead{$w_t$/[$E_{\text{net}},E_{\text{mem}},E_{\text{res}}$]} \\
\midrule
BackgroundImage & \makecell{$1.0$/[0.50, 0.70, 0.50]} & \makecell{$0.6$/[0.30, 0.70, 0.40]} & \makecell{$0.7$/[0.30, 0.70, 0.30]} & \makecell{$0.5$/[0.60, 0.90, 0.10]} & \makecell{$0.8$/[0.30, 0.80, 0.20]} & \makecell{$0.67$/[0.30, 0.70, 0.30]} \\
Bottom\_Navigation & \makecell{$0.8$/[0.50, 0.30, 0.40]} & \makecell{$0.3$/[0.20, 0.10, 0.30]} & \makecell{$0.4$/[0.10, 0.30, 0.50]} & \makecell{$1.1$/[0.80, 0.40, 0.50]} & \makecell{$0.3$/[0.00, 0.20, 0.40]} & \makecell{$0.46$/[0.30, 0.20, 0.40]} \\
Card & \makecell{$0.7$/[0.40, 0.30, 0.50]} & \makecell{$0.4$/[0.40, 0.30, 0.20]} & \makecell{$0.5$/[0.20, 0.60, 0.40]} & \makecell{$0.6$/[0.60, 0.50, 0.40]} & \makecell{$0.4$/[0.00, 0.30, 0.20]} & \makecell{$0.48$/[0.33, 0.30, 0.34]} \\
CheckBox(box) & \makecell{$0.2$/[0.10, 0.10, 0.20]} & \makecell{$0.1$/[0.10, 0.00, 0.10]} & \makecell{$0.1$/[0.00, 0.10, 0.10]} & \makecell{$0.2$/[0.30, 0.10, 0.10]} & \makecell{$0.1$/[0.00, 0.00, 0.10]} & \makecell{$0.12$/[0.10, 0.10, 0.10]} \\
CheckedTextView & \makecell{$0.3$/[0.20, 0.10, 0.30]} & \makecell{$0.1$/[0.10, 0.00, 0.10]} & \makecell{$0.2$/[0.00, 0.20, 0.30]} & \makecell{$0.1$/[0.10, 0.10, 0.10]} & \makecell{$0.2$/[0.00, 0.10, 0.20]} & \makecell{$0.16$/[0.10, 0.10, 0.20]} \\
Drawer & \makecell{$0.7$/[0.30, 0.20, 0.30]} & \makecell{$0.5$/[0.30, 0.20, 0.40]} & \makecell{$0.6$/[0.30, 0.40, 0.60]} & \makecell{$0.9$/[0.80, 0.40, 0.50]} & \makecell{$0.4$/[0.10, 0.30, 0.40]} & \makecell{$0.56$/[0.30, 0.30, 0.40]} \\
EditText & \makecell{$1.0$/[0.70, 0.30, 0.40]} & \makecell{$0.7$/[0.60, 0.10, 0.30]} & \makecell{$0.8$/[0.60, 0.20, 0.40]} & \makecell{$0.5$/[0.70, 0.10, 0.20]} & \makecell{$0.6$/[0.20, 0.10, 0.30]} & \makecell{$0.64$/[0.60, 0.20, 0.40]} \\
Icon & \makecell{$0.3$/[0.20, 0.30, 0.20]} & \makecell{$0.2$/[0.10, 0.20, 0.10]} & \makecell{$0.3$/[0.10, 0.30, 0.20]} & \makecell{$0.1$/[0.10, 0.20, 0.10]} & \makecell{$0.2$/[0.00, 0.20, 0.10]} & \makecell{$0.21$/[0.10, 0.20, 0.10]} \\
Image & \makecell{$1.2$/[0.70, 0.80, 0.40]} & \makecell{$0.8$/[0.70, 0.60, 0.20]} & \makecell{$1.2$/[0.70, 0.90, 0.50]} & \makecell{$1.2$/[0.90, 1.00, 0.20]} & \makecell{$1.0$/[0.50, 0.80, 0.30]} & \makecell{$1.01$/[0.70, 0.83, 0.32]} \\
Map & \makecell{$1.5$/[0.90, 0.90, 0.90]} & \makecell{$1.4$/[0.90, 0.80, 0.90]} & \makecell{$1.5$/[0.90, 1.00, 1.00]} & \makecell{$1.5$/[1.00, 0.90, 1.00]} & \makecell{$1.2$/[0.80, 0.90, 0.90]} & \makecell{$1.32$/[0.90, 0.90, 0.90]} \\
Modal & \makecell{$0.8$/[0.50, 0.20, 0.40]} & \makecell{$0.6$/[0.30, 0.30, 0.50]} & \makecell{$0.7$/[0.40, 0.50, 0.60]} & \makecell{$0.4$/[0.50, 0.20, 0.40]} & \makecell{$0.5$/[0.20, 0.30, 0.40]} & \makecell{$0.53$/[0.40, 0.30, 0.40]} \\
Multi\_Tab & \makecell{$1.0$/[0.60, 0.50, 0.60]} & \makecell{$0.7$/[0.50, 0.30, 0.40]} & \makecell{$0.9$/[0.50, 0.60, 0.80]} & \makecell{$1.0$/[0.80, 0.70, 0.50]} & \makecell{$0.6$/[0.30, 0.40, 0.60]} & \makecell{$0.78$/[0.60, 0.50, 0.60]} \\
PageIndicator & \makecell{$0.3$/[0.10, 0.10, 0.20]} & \makecell{$0.2$/[0.00, 0.10, 0.20]} & \makecell{$0.2$/[0.00, 0.10, 0.20]} & \makecell{$0.1$/[0.00, 0.10, 0.30]} & \makecell{$0.3$/[0.00, 0.10, 0.20]} & \makecell{$0.23$/[0.00, 0.10, 0.20]} \\
Remember & \makecell{$0.2$/[0.10, 0.05, 0.10]} & \makecell{$0.3$/[0.30, 0.20, 0.10]} & \makecell{$0.3$/[0.10, 0.10, 0.10]} & \makecell{$0.2$/[0.30, 0.10, 0.10]} & \makecell{$0.4$/[0.10, 0.20, 0.20]} & \makecell{$0.30$/[0.10, 0.10, 0.10]} \\
Spinner & \makecell{$0.9$/[0.60, 0.30, 0.30]} & \makecell{$0.5$/[0.40, 0.10, 0.20]} & \makecell{$0.9$/[0.50, 0.30, 0.40]} & \makecell{$0.5$/[0.80, 0.20, 0.20]} & \makecell{$0.7$/[0.20, 0.20, 0.40]} & \makecell{$0.64$/[0.60, 0.30, 0.40]} \\
Switch & \makecell{$0.6$/[0.30, 0.20, 0.40]} & \makecell{$0.1$/[0.10, 0.00, 0.10]} & \makecell{$0.2$/[0.00, 0.10, 0.10]} & \makecell{$0.2$/[0.30, 0.10, 0.20]} & \makecell{$0.2$/[0.00, 0.00, 0.10]} & \makecell{$0.28$/[0.14, 0.10, 0.13]} \\
Text & \makecell{$0.1$/[0.05, 0.05, 0.10]} & \makecell{$0.1$/[0.00, 0.00, 0.10]} & \makecell{$0.1$/[0.00, 0.00, 0.10]} & \makecell{$0.1$/[0.10, 0.10, 0.10]} & \makecell{$0.1$/[0.00, 0.00, 0.10]} & \makecell{$0.10$/[0.00, 0.00, 0.10]} \\
TextButton & \makecell{$0.6$/[0.50, 0.10, 0.30]} & \makecell{$0.4$/[0.40, 0.00, 0.20]} & \makecell{$0.3$/[0.10, 0.10, 0.20]} & \makecell{$0.7$/[0.90, 0.10, 0.20]} & \makecell{$0.3$/[0.00, 0.10, 0.20]} & \makecell{$0.43$/[0.38, 0.10, 0.20]} \\
Toolbar & \makecell{$0.4$/[0.10, 0.20, 0.20]} & \makecell{$0.3$/[0.20, 0.10, 0.20]} & \makecell{$0.6$/[0.20, 0.30, 0.50]} & \makecell{$0.3$/[0.30, 0.20, 0.30]} & \makecell{$0.5$/[0.10, 0.20, 0.40]} & \makecell{$0.39$/[0.18, 0.20, 0.32]} \\
UpperTaskBar & \makecell{$0.4$/[0.10, 0.20, 0.30]} & \makecell{$0.3$/[0.20, 0.10, 0.20]} & \makecell{$0.5$/[0.20, 0.30, 0.40]} & \makecell{$0.3$/[0.30, 0.20, 0.30]} & \makecell{$0.4$/[0.10, 0.20, 0.30]} & \makecell{$0.40$/[0.18, 0.20, 0.30]} \\
\bottomrule
\end{tabular}
}
\end{table*}

Table \ref{tab:model_comparison} captures a structured expert-elicitation exercise across five advanced LLMs under identical constraints. As expected, absolute values vary by model. Still, the rank-order and directional patterns are stable, meaning inputs that typically trigger network calls (e.g., EditText, Search/TextBox) skew toward higher $E_{net}$; bitmap-heavy elements (Image, BackgroundImage) load $E_{mem}$; render/interaction intensive widgets (Map, Modal, Drawer, Multi\_Tab) lean to $E_{res}$. In contrast, the UI-impact prior after robust aggregation and normalisation, the resultant $w_t$ and effect splits encode domain-plausible burdens, presenting the fact that Map and Image carry higher $w_t$ with memory/compute emphasis (tiles, bitmaps, WebView textures); EditText and TextButton retain modest weights but higher $E_{net}$ (type-ahead, auth, form posts); Modal/Drawer/Multi\_Tab tilt to $E_{res}$ (layout passes, animations), and purely textual widgets keep minimal weights. 

\begin{table}[!b]
\centering
\small
\caption{UI impact constants: per-axis soft caps $\kappa$ and sigmoid scales $\tau$.}
\label{tab:ui_axis_constants}
\vspace{2pt}
\begin{tabular}{lcc}
\toprule
Axis & Cap $\kappa_a$ & Scale $\tau_a$ \\
\midrule
\texttt{H} & 6.7310 & 1.3077 \\
\texttt{M}     & 6.7041 & 1.3147 \\
\texttt{R}  & 3.1326 & 0.8558 \\
\bottomrule
\end{tabular}
\end{table}

Table~\ref{tab:ui_axis_constants} sets per–axis soft caps $\kappa$ and sigmoid scales $\tau$ to regularise the translation from summed component impacts to bounded intent scores. These constants are empirically estimated on the RICO dataset using \uixpose{} deterministic pipeline as follow: (1) probe and build-evidence convert RICO annotations to per-screen EVIDENCE; (2) score aggregates component contributions with the Huber-prior weights/effects to produce pre-squash sums per axis; (3) stats derives caps $\kappa$ from the high tail of those sums ($H$: $\sim99^\text{th}$ percentile (pct); $M/R$: $\sim98^\text{th}$ pct) and sets $\tau$ so the $75^\text{th}$ percentile maps to $\approx0.8$ after the sigmoid. The learned caps ($\kappa$)
%($\kappa_{\text{net}}{=}6.7310$, $\kappa_{\text{mem}}{=}6.7041$, $\kappa_{\text{res}}{=}3.1326$)
prevent dense screens with many small elements from overwhelming the prior while still allowing high-impact components (e.g., \texttt{Map}, \texttt{EditText}, \texttt{TextButton}) to dominate when present. The calibrated scales ($\tau$) %($\tau_{\text{net}}{=}1.3077$, $\tau_{\text{mem}}{=}1.3147$, $\tau_{\text{res}}{=}0.8558$) 
yield a flatter response for network/memory (smoothing bursty fetches and steady allocations) and a steeper response for resources (earlier saturation under CPU/GPU-heavy UI). In practice, these corpus-derived constants stabilise scores across heterogeneous layouts, improve rank consistency for downstream alignment, and reduce double-counting. 

\subsubsection{H-Channel}
\label{subsubsec:h-channel}

We derive H-channel hyperparameters empirically using the \texttt{AndMal2020} dataset \cite{keyes2021entroplyzer} to ground the detection logic in historical malware behaviour. The initialisation process consists of three stages.

We separate features into \textit{physical deltas}, i.e., actual system stress, and \textit{risk surrogates} as behavioural indicators. Physical deltas quantify network throughput (bytes/packets) and host memory pressure (PSS/Heap). From API calls and logs, we construct six rank-normalised surrogate scores, which are \textit{requests} (session/WebView volume), \textit{failures} (error logs), \textit{cross-domain} (risky WebView navigation), \textit{reputation} (crypto/obfuscation primitives), \textit{trackers} (identifier leakage), and \textit{malicious-bytes} (execution signals). These are mapped to $[0,1]$ via percentile ranking to ensure comparability across heterogeneous scales.

We measure the predictive power of each surrogate against physical deltas using a blended magnitude of Spearman (rank) and Pearson (linear) correlations. To control for false discoveries across the effect grid, we apply the Benjamini–Hochberg procedure, shrinking any surrogate that fails to yield statistically significant associations with system pressure. A baseline score for raw traffic volume is computed similarly to anchor the semantics to observed traffic magnitudes.

Final H-channel weights are obtained by normalising these empirical correlation strengths. To prevent coincident coupling between memory and network activity from skewing the results, we compute the average absolute cross-correlation between host memory and throughput deltas. This value is transformed into small, capped spillover terms that regularise the channel fusion.

\subsubsection{M-Channel} \label{subsubsec:m-channel} We derive M-channel hyperparameters by analysing memory dynamics within the paired delta table. Primary components include rate-based metrics, e.g., PSS and heap allocation deltas, and saturation indicators, i.e., swap growth and heap pressure, computed via post-state or delta ratios. We quantify the alignment of these components with network throughput deltas using the exact robust correlation and shrinkage procedure described in the H-channel. The resulting strengths are normalised into a memory mixture, subject to a soft design prior that ensures rate-based components retain at least as much influence as static level indicators via rebalancing.

To capture network-adjacent overhead, we introduce lightweight ``handshake'' proxies, such as binder/parcel growth and WebView count growth. These are assigned weights proportional to their empirical strength against throughput, capped at a fixed total budget to prevent them from overshadowing byte-level evidence; components lacking significant association are shrunk to zero. Similarly, UI churn is estimated from view and root-view deltas relative to host memory pressure and clipped to a modest cap to prevent signal amplification on benign, UI-dense screens.

Finally, to distinguish sustained pressure from transient noise, the composite memory score is gated at a high empirical quantile. Only observations exceeding this threshold contribute to the accumulator, and a short EWMA window is applied to smooth the signal without masking rapid state changes.

\subsubsection{R-Channel} \label{subsubsec:r-channel} The R-channel is similarly derived from the paired delta table; however, due to the absence of direct resource metrics (e.g., \texttt{CPU\%}, core counts, or scheduler fields) in the reference dataset, standard compute magnitude and volatility terms are disabled. Instead, the channel functions as a scheduler-side corroboration layer, relying solely on memory-visible dynamics that change in proportional set size (PSS), heap allocation, swap dirty pages, and heap pressure. Each component is rank-normalised to ensure comparability.

We measure the empirical association of each corroborator against \textit{host memory-pressure deltas}, applying the same blended correlation procedure and per-feature false discovery control used elsewhere. Components lacking statistical support are shrunk to zero; the remaining significant strengths are normalised into a corroborator mixture. The system handles missing data dynamically; weights are renormalised based on the available subset of features, reverting to a default distribution only if all candidates are absent.

To prioritise sustained resource intensity over incidental churn, the final mixture is gated by a high empirical quantile derived from the dataset's observed distribution and smoothed with a short exponentially weighted window. Consequently, the R-channel contributes only when scheduler-visible dynamics persistently elevate.

Finally, \uixpose{} considers fusion weights ($\alpha_k$ in Equation \ref{eq:behfusion}) with the default $\alpha_H{=}0.4$, $\alpha_M{=}\alpha_R{=}0.3$ to aggregate the behaviour vector before performing alignment analysis. 

The resulting prior $\omega_t$ and $E_t$  and per-axis caps/scales $\kappa_a$ and $\tau_a$ are used unchanged in all subsequent experiments (Sections \ref{sec:experiment_res} and \ref{sec:casestudies_res}), i.e., the Wikipedia, SmartCurrencyConverter, and Flashy case studies. 

\subsection{Experimental Setup \& UI Automation}
\label{sec:experiment_res}
For the experimental setup, \uixpose{} employs either a smartphone or the Android emulator source code (version $35.1.13.0$)\footnote{https://android.googlesource.com/platform/external/qemu/+/emu-master-dev}, modified to support traffic flow capturing on the fly by libpcap-based capture scripts embedded in the emulator to extract packet features. The emulator utilises a packet queue with a customisable length ($l=5$ in this study) that determines the frequency of flow feature updates. The quick boot feature is also disabled on the emulator to prevent reloading previously saved snapshots. \uixpose{} employs a modified version of DroidBot as the UI automation engine, configured to automatically generate UI interaction sequences to explore each app, while \uixpose{} collects runtime semantics. \uixpose{} runs each application for 30 minutes (spanning 5-second steps), and each captured UI is timestamped for processing.
%, hash-based queuing system to buffer unprocessed screens and store the corresponding hash, thereby avoiding reprocessing in the future. 

\subsection{Results}
\label{sec:casestudies_res}
With the alignment prior in place, we next characterise the practical cost of intent inference by measuring LLM processing delay.

Finally, using the same prior and experimental setup, we apply \uixpose{} to three real-world Android apps to qualitatively assess the kinds of anomalies it surfaces.

\subsubsection{LLMs Processing Delay}

We measure the end-to-end delay to obtain the intention vector $i_t$ for a given UI by issuing a series of requests to a vision-based LLM (ChatGPT). For each screen sampled at random from the RICO dataset, we execute Algorithm~\ref{alg:ui_to_evidence} and record server-reported token-usage and streaming-latency milestones. As Algorithm~\ref{alg:evidence_to_intent} is deterministic, it is omitted from the evaluation. 

For every request, we read the API usage block to obtain exact token counts; \texttt{completion\_tokens} ($ct$) and \texttt{total\_tokens} ($tt$). 
These values include \emph{vision} tokens for the screenshot as billed by the provider. We also report throughput during generation as
$\text{TPS}^{(s)} = \frac{\texttt{ct}^{(s)}}{\tau_{\text{gen}}^{(s)}} \quad \text{(tokens per second)}$.

For each UI, we store a compact record along with UI-structure features derived from Algorithm~\ref{alg:ui_to_evidence}, i.e., component counts, weighted complexity, actionable controls. This avoids unstable proxies such as pixel density or downscaling size, since vision tokenisation is already reflected in \texttt{prompt\_tokens}. We report the mean and dispersion of TPS across the sampled RICO screens as the LLM processing-delay profile for $i_t$ extraction. In the OpenAI API, these hyperparameters (temperature = 0.2 and detail = high) are used in our experiment, applying only when an image is included. They instruct the vision-capable model to process images with higher fidelity, including detailed OCR and finer features. Temperature controls randomness in text generation, meaning lower values produce more deterministic, focused outputs, ideal for math or accuracy tasks.

\begin{figure}[!t]
    \centering
    \begin{minipage}{0.49\columnwidth}
        \centering
        \includegraphics[keepaspectratio, scale=0.38, trim={0 0 0 0}, clip]{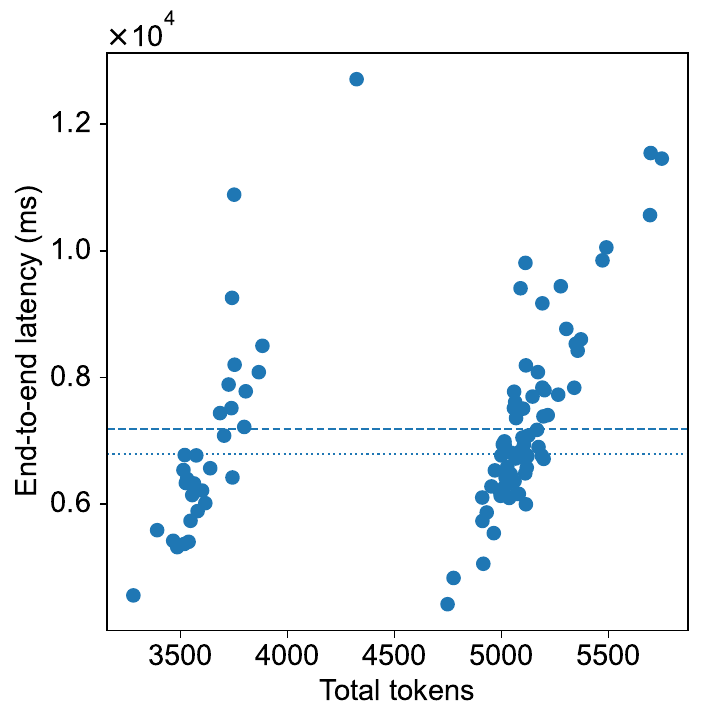}
        \subcaption{}
        \label{fig:tt_lat}
    \end{minipage}
   \hfill
   \begin{minipage}{0.49\columnwidth}
        \centering
        \includegraphics[keepaspectratio, scale=0.38, trim={0 0 0 0}, clip]{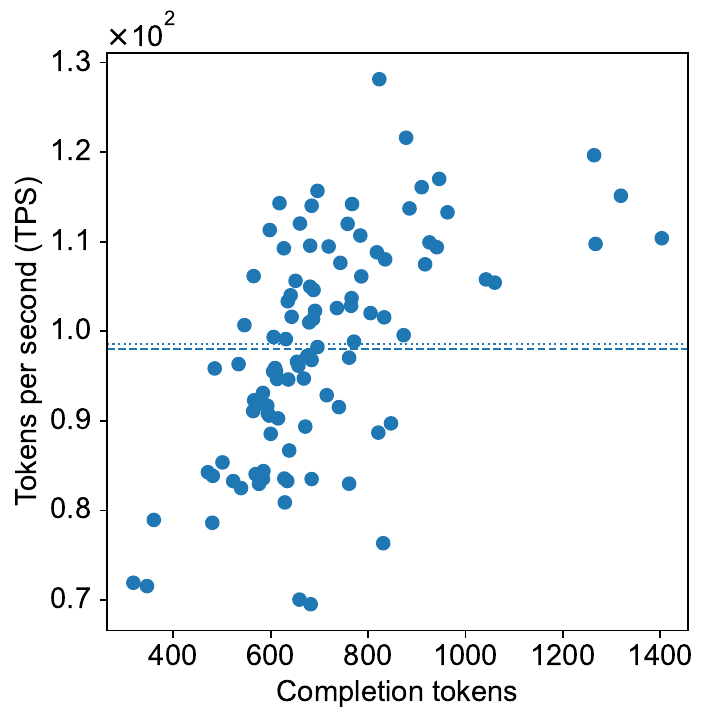}
        \subcaption{}
        \label{fig:rct_tps}
    \end{minipage}
    \begin{minipage}{0.49\columnwidth}
        \centering
        \includegraphics[keepaspectratio, scale=0.38, trim={0 0 0 0}, clip]{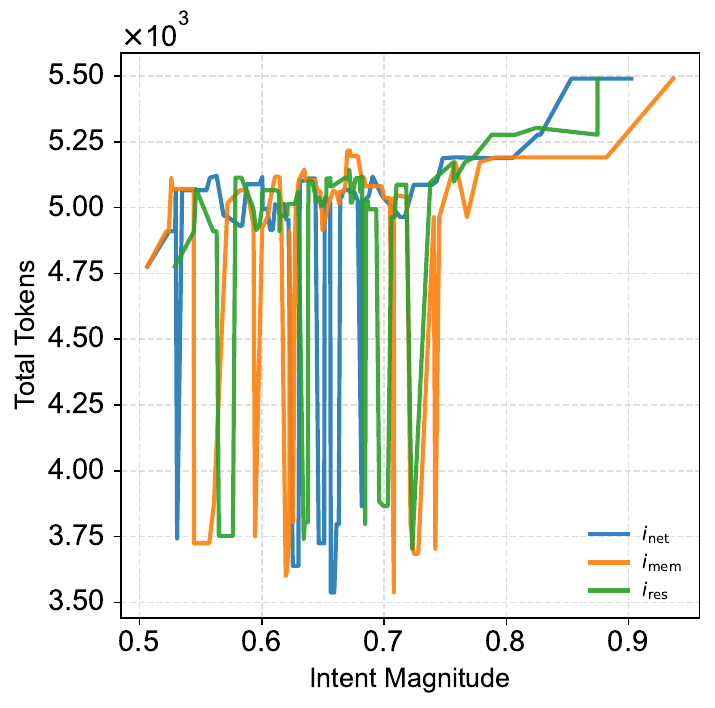}
        \subcaption{}
        \label{fig:i_tt}
    \end{minipage}
   \hfill
   \begin{minipage}{0.49\columnwidth}
        \centering
        \includegraphics[keepaspectratio, scale=0.38, trim={0 0 0 0}, clip]{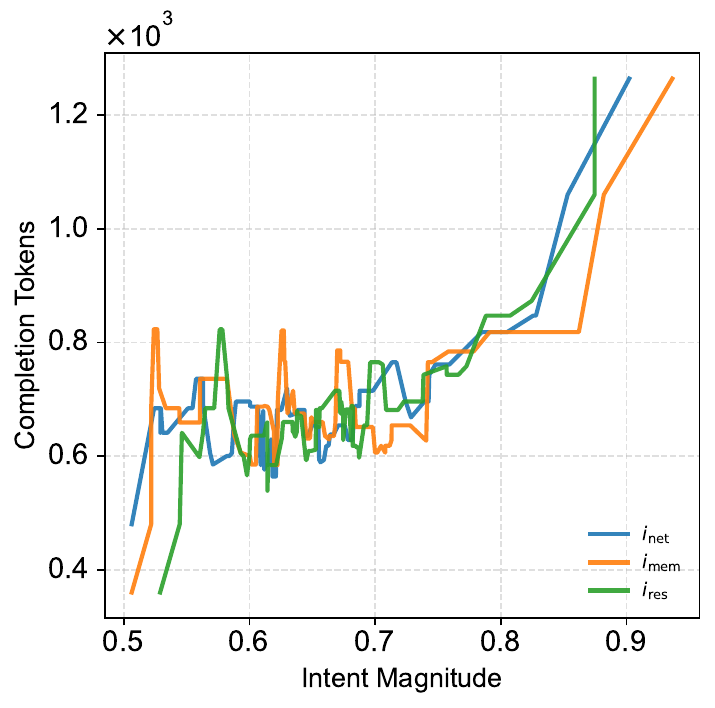}
        \subcaption{}
        \label{fig:i_ct}
    \end{minipage}
    \caption{Token-performance and intent scaling (a) latency vs. total tokens; (b) throughput (tokens/s) vs. completion tokens (means dashed); (c) total tokens vs. intent magnitude; (d) completion tokens vs. intent magnitude.}
    \label{fig:res_pred}
\label{fig:llm_delay}
\end{figure}

Figure \ref{fig:llm_delay}(a,b) presents the latency vs total tokens and TPS vs completion tokens over 100 randomly chosen UIs from the RICO dataset, with respect to the algorithm's processing.  Figure \ref{fig:tt_lat} shows a positive relationship between request size and end-to-end latency. The mean and median reference lines suggest a slightly right-skewed distribution, consistent with occasional long-tail runs. Two distinct groups are visible; a lower-latency cluster centred around 5.0–5.3k tokens and a higher-latency cluster beyond 5.4k tokens, indicating heteroscedasticity and possibly hidden factors such as prompt types or server/network conditions. The overall upward trend supports size-driven costs, and although a few outliers remain influential. Figure \ref{fig:rct_tps} shows that generation throughput remains stable across $3.5 \times 10^2$ to $1.3 \times 10^3$ completion tokens, with minor variation. The mean and median guides indicate that TPS dependence on output length is weak; any trend is minor compared to the dispersion. 

Figure \ref{fig:llm_delay}(c,d) illustrates the aggregate behavioural coupling. Each intent component represents a resource-axis projection derived from static RICO UI samples. When mapped against the measured token-generation metrics of Algorithm 1, the resulting trajectories reveal how visual-compositional load translates into a quantifiable modulation of model throughput.

In Figure \ref{fig:i_tt} all three trends exhibit a stable but compressive regime across mid-range intent magnitudes (0.55–0.8), followed by a divergence at higher values. The upward inflection in total tokens suggests that screens with compounded network or memory cues trigger longer completions, consistent with semantically richer prompts generated by Algorithm 1’s evidence synthesis. However, the convergence of $i_{mem}$ and $i_res$ at the high end implies a shared saturation limit—beyond which additional component density no longer scales output length linearly. This behaviour reflects a bounded elaboration effect, as the model’s autoregressive generator exhibits diminishing returns when the visual entropy exceeds an internal contextual threshold.
Figure \ref{fig:i_ct} demonstrates that completion tokens follow a similar rising trajectory but with greater noise sensitivity at lower magnitudes, implying that even small increments in perceived UI complexity can perturb early decoding phases. The near-parallel slopes of $i_{net}$ and $i_{res}$ indicate that both network-heavy and computation-heavy interfaces bias the LLM toward verbosity. At the same time, memory-dominant screens yield more tempered but consistent completions. 

\subsubsection{Applications}

We built the test applications dataset using information from major security blogs such as McAfee and Kaspersky. The list of application hashes is available on the \uixpose{} repository. Using the prior and hyperparameters from Section \ref{sec:ui_impact_hyperparameter}, we now apply \uixpose{} to three representative Android apps, guided by the following questions:

\textbf{RQ1 - Alignment on benign apps / false positives.} When applied to benign apps, does \uixpose{} classify visually justified, resource-heavy states as authorised, thereby avoiding systematic false positives on normal behaviour?

\textbf{RQ2 - Localising misbehaviour / false negatives.} For a known-malicious app, can \uixpose{} highlight low-alignment, high-magnitude states that correspond to covert or undesired behaviour, rather than remaining silent (false negatives)?

\textbf{RQ3 - Distinguishing benign-but-buggy from malicious.} For a benign but unstable app, do \uixpose{}’s anomalies align with implementation issues (e.g., leaks, crash-related spikes) rather than with exfiltration-like patterns, helping distinguish buggy utilities from truly malicious apps?

\paragraph{Wikipedia as Benign Case Study \cite{wikipedia}}

To illustrate how \uixpose{} combines visual intent and runtime behaviour, we ran the open-source Wikipedia Android app in the instrumented emulator, including app launch, home/Explore navigation, text search, voice search, and the ``Saved pages'' feature. For each frame, sampled every five steps up to step~300, the LLM predicted an \emph{intent vector} over network, memory, and CPU usage, which we compared with the normalised behaviour referred to as \emph{observed} channel values.

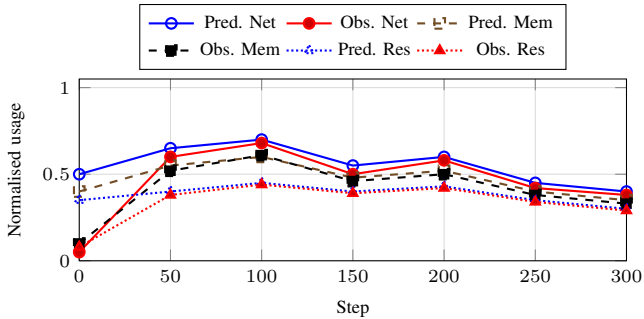
\begin{figure}[!t]
  \centering
  \pgfplotstableread[row sep=\\,col sep=space]{
    step   p_net   o_net   p_mem   o_mem   p_cpu   o_cpu \\
    0      0.50    0.05    0.40    0.10    0.35    0.08  \\
    50     0.65    0.60    0.55    0.52    0.40    0.38  \\
    100    0.70    0.68    0.60    0.61    0.45    0.44  \\
    150    0.55    0.50    0.48    0.46    0.40    0.39  \\
    200    0.60    0.58    0.52    0.50    0.43    0.42  \\
    250    0.45    0.42    0.40    0.38    0.35    0.34  \\
    300    0.40    0.38    0.35    0.33    0.30    0.29  \\
  }\wikipediadata

  \begin{tikzpicture}
    \begin{axis}[
      width=1\columnwidth,
      height=4.0cm,
      xlabel={Step},
      ylabel={Normalised usage},
      xmin=0, xmax=300,
      ymin=0, ymax=1.05,
      legend style={font=\scriptsize, at={(0.5,1.05)},anchor=south,legend columns=3},
      tick label style={font=\scriptsize},
      label style={font=\scriptsize},
      grid=both,
      major grid style={line width=.1pt,draw=gray!30},
      minor grid style={line width=.1pt,draw=gray!15},
    ]

      % Network
      \addplot+[mark=o,thick] table[x=step,y=p_net] {\wikipediadata};
      \addlegendentry{Pred.\ Net};
      \addplot+[mark=*,thick] table[x=step,y=o_net] {\wikipediadata};
      \addlegendentry{Obs.\ Net};

      % Memory
      \addplot+[mark=square,thick,dashed] table[x=step,y=p_mem] {\wikipediadata};
      \addlegendentry{Pred.\ Mem};
      \addplot+[mark=square*,thick,dashed] table[x=step,y=o_mem] {\wikipediadata};
      \addlegendentry{Obs.\ Mem};

      % CPU / Resource
      \addplot+[mark=triangle,thick,densely dotted] table[x=step,y=p_cpu] {\wikipediadata};
      \addlegendentry{Pred.\ Res};
      \addplot+[mark=triangle*,thick,densely dotted] table[x=step,y=o_cpu] {\wikipediadata};
      \addlegendentry{Obs.\ Res};

    \end{axis}
  \end{tikzpicture}
  \caption{Wikipedia case study. Predicted vs.\ observed Network / Memory / Resource across the UI trace.}
  \label{fig:wikipedia-timeline}
\end{figure}

Figure~\ref{fig:wikipedia-timeline} plots representative predicted and observed resource usage over time (steps 0--300). Phases with inherently heavy computation, such as Google voice search (around steps~95 and~195) and Explore content loading (around step~315, beyond this window), exhibit consistently high network and CPU usage. \uixpose{} correctly classifies these as \emph{Authorised}, even when the observed usage exceeds the predicted intent, because the UI clearly justifies the resource footprint (e.g., active microphone dialogue or loading spinner).

In contrast, \uixpose{} flags long sequences of visually simple or empty screens as anomalous, such as the ``Saved pages'' view reporting ``No saved pages yet'' and empty Explore feeds, where memory and CPU remain close to their maximum normalised values despite an idle or near-blank UI. These cases correspond to anomalies in which a static foreground UI does not justify sustained background activity (see representative examples in Table~\ref{tab:wiki-frames}).
%Finally, we observe a third class of discrepancy where the UI and behaviour agree. Still, the intent model over-predicts usage (e.g., an idle home screen with zero activity but medium predicted usage). We treat these as \emph{intent model errors}, rather than security-relevant anomalies, which helps separate backend misbehaviour from LLM calibration issues in our analysis (Table~\ref{tab:wiki-ui-classes}).

\begin{table}[!t]
  \centering
  \caption{Representative frames from the Wikipedia case study with normalised values in $[0,1]$ for \textbf{N}etwork, \textbf{M}emory, \textbf{R}esource.
  % The last row illustrates an \emph{intent-model} anomaly: UI and behaviour match, but the predicted intent is overly high.
  }
  \label{tab:wiki-frames}
  \scriptsize
  \begin{tabular}{@{}r p{2.3cm} p{1.8cm} p{1.9cm} p{1.4cm}@{}}
    \toprule
    Step & UI state & Intent (N/M/R) & Observed (N/M/R) & Verdict \\
    \midrule
    95  & Google voice search dialogue (active speech input) &
          (0.56, 0.55, 0.60) &
          (0.68, 0.66, 0.66) &
          Authorised \\
    225 & Empty ``Saved pages'' list (``No saved pages yet'') &
          (0.54, 0.53, 0.57) &
          (0.40, 0.50, 1.00) &
          Anomaly \\
    265 & Empty Explore feed (``There's nothing on your Explore feed'') &
          (0.56, 0.55, 0.61) &
          (0.50, 1.00, 1.00) &
          Anomaly \\
    % 15  & Idle Android home screen (launcher) &
    %       (0.54, 0.53, 0.58) &
    %       (0.00, 0.00, 0.00) &
    %       Anomaly (intent) \\
    \bottomrule
  \end{tabular}
\end{table}

\begin{table}[!b]
  \centering
  \caption{High-level breakdown of frames (up to step 300) by UI class and verdict. Counts are illustrative for this case study; the key trend is that voice search and loading UIs are almost always authorised, while simple empty lists and feeds trigger most anomalies.}
  \label{tab:wiki-ui-classes}
  \scriptsize
  \begin{tabular}{@{}r p{1.8cm} p{1.5cm} p{1.5cm} p{1.1cm}@{}}
    \toprule
    UI class                          & Authorised & Anomaly & Uncertain \\
    \midrule
    Launcher / Home                   & 3          & 1       & 0         \\
    Explore (loading / content)       & 8          & 3       & 1         \\
    Text search input / history       & 10         & 4       & 1         \\
    Voice search dialog               & 14         & 1       & 0         \\
    Empty Saved pages / lists         & 6          & 10      & 0         \\
    Settings / language management    & 4          & 5       & 0         \\
    \midrule
    Total                             & 45         & 24      & 2         \\
    \bottomrule
  \end{tabular}
\end{table}

\paragraph{SmartCurrencyConverter as Malicious Case Study}

We further evaluated \uixpose{} on \texttt{SmartCurrencyConverter}, a \emph{known malicious} third-party Android currency converter app. The interaction covers splash/startup, launcher, the main converter calculator, currency selection drawers, chart views (including error states), settings, and cross-promotion dialogues. Importantly, \uixpose{} is not given the malware label; it only sees the UI screenshots, the normalised network, memory, and resource traces, and the LLM-predicted intent vectors.

\begin{figure}[!t]
  \centering
  \pgfplotstableread[row sep=\\,col sep=space]{
    step   p_net   o_net   p_mem   o_mem   p_cpu   o_cpu \\
    0      0.604   0.000   0.583   0.000   0.682   0.000 \\
    45     0.567   0.358   0.554   0.000   0.621   0.000 \\
    60     0.568   0.429   0.554   0.638   0.621   0.000 \\
    70     0.567   0.538   0.554   1.000   0.620   0.956 \\
    135    0.539   0.470   0.531   0.530   0.570   0.969 \\
    150    0.564   0.391   0.551   1.000   0.614   0.401 \\
    215    0.584   0.555   0.568   0.850   0.650   1.000 \\
    245    0.580   0.332   0.564   0.334   0.643   1.000 \\
    295    0.553   0.260   0.543   0.426   0.596   0.957 \\
    300    0.548   0.231   0.538   0.496   0.586   0.745 \\
  }\currencydata

  \begin{tikzpicture}
    \begin{axis}[
      width=1\columnwidth,
      height=4.0cm,
      xlabel={Step},
      ylabel={Normalized usage},
      xmin=0, xmax=300,
      ymin=0, ymax=1.05,
      legend style={font=\scriptsize, at={(0.5,1.05)},anchor=south,legend columns=3},
      tick label style={font=\scriptsize},
      label style={font=\scriptsize},
      grid=both,
      major grid style={line width=.1pt,draw=gray!30},
      minor grid style={line width=.1pt,draw=gray!15},
    ]

      % Network
      \addplot+[mark=o,thick] table[x=step,y=p_net] {\currencydata};
      \addlegendentry{Pred.\ Net};
      \addplot+[mark=*,thick] table[x=step,y=o_net] {\currencydata};
      \addlegendentry{Obs.\ Net};

      % Memory
      \addplot+[mark=square,thick,dashed] table[x=step,y=p_mem] {\currencydata};
      \addlegendentry{Pred.\ Mem};
      \addplot+[mark=square*,thick,dashed] table[x=step,y=o_mem] {\currencydata};
      \addlegendentry{Obs.\ Mem};

      % CPU / Resource
      \addplot+[mark=triangle,thick,densely dotted] table[x=step,y=p_cpu] {\currencydata};
      \addlegendentry{Pred.\ Resource};
      \addplot+[mark=triangle*,thick,densely dotted] table[x=step,y=o_cpu] {\currencydata};
      \addlegendentry{Obs.\ Resource};

    \end{axis}
  \end{tikzpicture}
  \caption{SmartCurrencyConverter case study. Predicted vs.\ observed Network / Memory / Resource for representative steps. Note the sustained high resource and memory usage hidden behind visually simple calculator and dialogue screens, and the strong authorised high-network pattern consistent with a data-hungry but malicious currency app.}
  \label{fig:currencyconverter-timeline}
\end{figure}
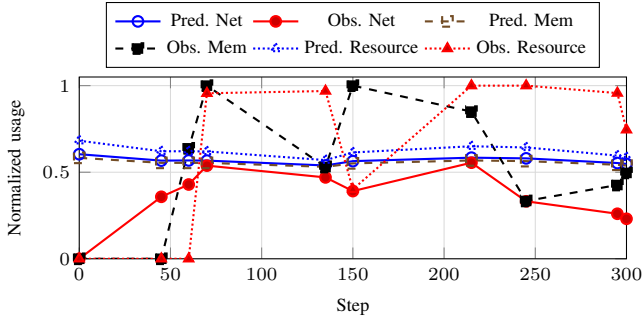

Figure~\ref{fig:currencyconverter-timeline} shows representative predicted and observed memory/resource usage over time (steps 0--300). For visually complex, data-heavy screens such as the main converter and currency selection drawers (e.g., around steps~60, 100, 130), \uixpose{} typically classifies medium-to-high resource usage as \emph{Authorised}, since this behaviour matches the app's apparent function, i.e., fetching exchange rates and rendering long lists of flags and currencies. In contrast, \uixpose{} flags as anomalous several visually simple or error states where resource usage spikes, particularly, static calculator screens with no visible interaction but very high resource (e.g., step~135 and step~155), chart error screens with maximum memory despite a prominent ``Chart failed'' message (step~150), and static recommendation dialogs that merely list other apps while resource is close to~1.0 (steps~215, 245, 285--295). These anomalies correspond to UI states where the app appears idle or in an error state, yet performs sustained heavy computation, a pattern consistent with hidden malicious routines, and in this case, multiple attempts to connect to a malicious endpoint, as ad/analytics abuse rather than user-visible conversion logic. 
%Early splash screens (steps~0 and~5) exhibit the opposite pattern: the intent model predicts medium-high activity, but the app is entirely idle; we treat these primarily as \emph{intent-model anomalies} rather than security-relevant misbehaviour. 
Table~\ref{tab:smart-frames} highlights representative frames, and Table~\ref{tab:smart-ui-classes} shows how anomalies cluster in non-core UI classes (chart errors, settings, recommendation dialogues), supporting \uixpose{}'s ability to localise suspicious behaviour within a known-malicious app.

\begin{table}[t]
  \centering
  \caption{Representative frames from SmartCurrencyConverter with normalised values in $[0,1]$ for \textbf{N}etwork, \textbf{M}emory, \textbf{R}esource. Anomalies align with visually simple or error-prone UIs that mask heavy computation.}
  \label{tab:smart-frames}
  \scriptsize
  \begin{tabular}{@{}r p{2.3cm} p{1.9cm} p{1.9cm} p{1.4cm}@{}}
    \toprule
    Step & UI state & Intent (N/M/R) & Observed (N/M/R) & Verdict \\
    \midrule
    60  & Main converter calculator (flags + keypad) &
          (0.57, 0.55, 0.62) &
          (0.43, 0.64, 0.00) &
          Authorised \\
    100 & Main converter with rates and ``TRENDS \& CHARTS'' button &
          (0.53, 0.53, 0.56) &
          (0.75, 0.31, 0.62) &
          Authorised \\
    150 & Chart error (``Chart failed'') &
          (0.56, 0.55, 0.61) &
          (0.39, 1.00, 0.40) &
          Anomaly (memory) \\
    215 & Currency list + error message in drawer &
          (0.58, 0.57, 0.65) &
          (0.56, 0.85, 1.00) &
          Anomaly (resource) \\
    \bottomrule
  \end{tabular}
\end{table}

\begin{table}[!t]
  \centering
  \caption{High-level breakdown of frames (up to step 300) for SmartCurrencyConverter (known-malicious sample) by UI class and verdict. Anomalies concentrate in non-core UI classes (chart-error screens, settings, recommendation dialogues), where the app appears idle but exhibits heavy background activity.}
  \label{tab:smart-ui-classes}
  \scriptsize
  \begin{tabular}{@{}r p{1.5cm} p{1.1cm} p{1.1cm} p{1.1cm}@{}}
    \toprule
    UI class                                    & Authorised & Anomaly & Uncertain \\
    \midrule
    Splash / Startup                            & 0          & 2       & 0         \\
    Launcher / Home                             & 7          & 0       & 0         \\
    Main converter / calculator                 & 12         & 5       & 0         \\
    Currency selection lists (flags + names)    & 8          & 2       & 0         \\
    Currency charts (incl.\ error screens)      & 9          & 3       & 0         \\
    Settings screens                            & 2          & 1       & 0         \\
    App recommendation / cross-promo dialogs    & 6          & 4       & 0         \\
    \midrule
    Total                                       & 44         & 17      & 0         \\
    \bottomrule
  \end{tabular}
\end{table}

\paragraph{Flashy Open-Source F-Droid App as Benign Case Study}

As a benign reference point, we apply \uixpose{} to \texttt{Flashy}, an open-source flashlight utility distributed via F-Droid. The UI trace covers device boot, launcher navigation, the main flashlight control panel (power/mode/SOS/brightness/colour wheel), settings and theme dialogues, Android crash dialogues, system \emph{App info} screens, and the app’s \emph{About} page. Again, \uixpose{} only sees screenshots, per-step (Network/Memory/Resource) traces, and LLM-predicted intent vectors; it does \emph{not} use the prior knowledge that the app is benign.

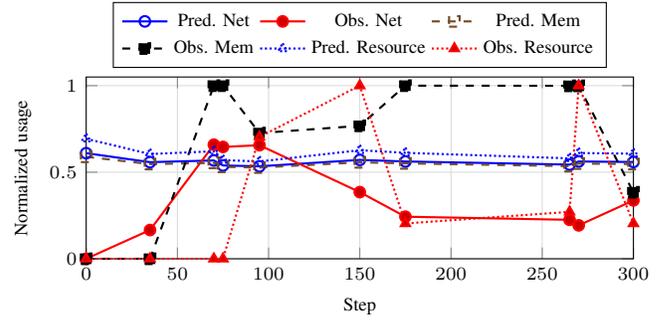
\begin{figure}[!t]
  \centering
  \pgfplotstableread[row sep=\\,col sep=space]{
    step   p_net   o_net   p_mem   o_mem   p_cpu   o_cpu \\
    0      0.610   0.000   0.589   0.000   0.693   0.000 \\
    35     0.558   0.166   0.547   0.000   0.605   0.000 \\
    70     0.568   0.659   0.554   1.000   0.621   0.000 \\
    75     0.539   0.646   0.531   1.000   0.570   0.000 \\
    95     0.534   0.656   0.527   0.726   0.562   0.704 \\
    150    0.571   0.385   0.557   0.768   0.627   1.000 \\
    175    0.562   0.243   0.551   1.000   0.612   0.204 \\
    265    0.544   0.225   0.535   0.999   0.580   0.271 \\
    270    0.562   0.193   0.550   0.999   0.611   1.000 \\
    300    0.559   0.337   0.547   0.384   0.607   0.204 \\
  }\flashydata

  \begin{tikzpicture}
    \begin{axis}[
      width=1\columnwidth,
      height=4.0cm,
      xlabel={Step},
      ylabel={Normalized usage},
      xmin=0, xmax=300,
      ymin=0, ymax=1.05,
      legend style={font=\scriptsize, at={(0.5,1.05)},anchor=south,legend columns=3},
      tick label style={font=\scriptsize},
      label style={font=\scriptsize},
      grid=both,
      major grid style={line width=.1pt,draw=gray!30},
      minor grid style={line width=.1pt,draw=gray!15},
    ]

      % Network
      \addplot+[mark=o,thick] table[x=step,y=p_net] {\flashydata};
      \addlegendentry{Pred.\ Net};
      \addplot+[mark=*,thick] table[x=step,y=o_net] {\flashydata};
      \addlegendentry{Obs.\ Net};

      % Memory
      \addplot+[mark=square,thick,dashed] table[x=step,y=p_mem] {\flashydata};
      \addlegendentry{Pred.\ Mem};
      \addplot+[mark=square*,thick,dashed] table[x=step,y=o_mem] {\flashydata};
      \addlegendentry{Obs.\ Mem};

      % CPU / Resource
      \addplot+[mark=triangle,thick,densely dotted] table[x=step,y=p_cpu] {\flashydata};
      \addlegendentry{Pred.\ Resource};
      \addplot+[mark=triangle*,thick,densely dotted] table[x=step,y=o_cpu] {\flashydata};
      \addlegendentry{Obs.\ Resource};

    \end{axis}
  \end{tikzpicture}
  \caption{Flashy case study. Predicted vs.\ observed Network / Memory / Resource. Unlike the known malicious currency app, most authorised frames show modest, well-aligned network usage, with anomalies dominated by memory/resource spikes in settings and crash-related states.}
  \label{fig:flashy-timeline}
\end{figure}

Figure~\ref{fig:flashy-timeline} summarises representative steps between 0 and~300. As expected, most launcher and control-panel frames are classified as \emph{Authorised}: network and memory usage are low-to-me, and this is and visually justified by the colour wheel and interactive controls (e.g., steps~130–140, 170, 195, 205--215, 230--240, 290--310). In contrast, \uixpose{} flags as anomalous several clusters of frames where a visually simple or even blank UI exhibits very high resource usage, e.g., settings and theme-selection dialogues with near-maximal memory or resource (steps~70--80, 85--105, 110--120), static home or control screens with unexpectedly high memory (steps~145, 175, 185, 190, 220, 225), and crash/\emph{App info} states or a black screen that still consume extreme memory or resource (steps~155, 180, 190, 260--270, 275). Table~\ref{tab:flashy-frames} highlights a subset of these frames, and Table~\ref{tab:flashy-ui-classes} aggregates them by UI class. 

Crucially, unlike the known-malicious \texttt{SmartCurrencyConverter}, these anomalies in \texttt{Flashy} \emph{do not} line up with stealthy background activity hidden behind apparently legitimate, data-heavy views. Instead, they mostly occur in settings, theme dialogues, and crash/\emph{App info} states, and often form short-lived bursts around instability events. Given that \texttt{Flashy} is small, open-source and non-networked beyond basic analytics, this pattern is more consistent with inefficient memory management or implementation bugs than with exfiltration or covert computation. In other words, \uixpose{} still surfaces useful ``red flags'', e.g., resource leaks, instability, even for benign apps, but the distribution of anomalies across UI classes helps distinguish buggy-but-benign utilities from truly suspicious, known-malicious behaviour such as that observed in \texttt{SmartCurrencyConverter}.

\begin{table}[!b]
  \centering
  \caption{Representative frames from Flashy app. Intent and observed values are normalised in $[0,1]$ for \textbf{N}etwork, \textbf{M}emory, \textbf{R}esource. Anomalies tend to coincide with settings, crash, or \emph{App info} states rather than the core control UI.}
  \label{tab:flashy-frames}
  \scriptsize
  \begin{tabular}{@{}r p{2.3cm} p{1.9cm} p{1.9cm} p{1.4cm}@{}}
    \toprule
    Step & UI state & Intent (N/M/R) & Observed (N/M/R) & Verdict \\
    \midrule
    35  & Flashy control panel (idle, colour wheel visible) &
           (0.56, 0.55, 0.60) &
           (0.17, 0.00, 0.00) &
           Authorised \\
    75  & Settings screen (static list of options) &
           (0.54, 0.53, 0.57) &
           (0.65, 1.00, 0.00) &
           Anomaly (memory) \\
    150 & Flashy control panel (idle controls) &
           (0.57, 0.56, 0.63) &
           (0.38, 0.77, 1.00) &
           Anomaly (resource) \\
    180 & System \emph{App info} for Flashy &
           (0.54, 0.53, 0.57) &
           (0.27, 1.00, 0.20) &
           Anomaly (memory) \\
    270 & Flashy control panel after crash &
           (0.56, 0.55, 0.61) &
           (0.19, 1.00, 1.00) &
           Anomaly (mem+resource) \\
    300 & About screen for Flashy &
           (0.56, 0.55, 0.61) &
           (0.34, 0.38, 0.20) &
           Authorised \\
    \bottomrule
  \end{tabular}
\end{table}

\begin{table}[!t]
  \centering
  \caption{High-level breakdown of frames (up to step 300) for Flashy by UI class and verdict. Anomalies concentrate in settings, theme dialogues, crash/\emph{App info} states, and a few control-panel frames around crashes, consistent with resource leaks or bugs rather than covert malicious behaviour.}
  \label{tab:flashy-ui-classes}
  \scriptsize
  \begin{tabular}{@{}r p{1.4cm} p{1.4cm} p{1.4cm} p{1.1cm}@{}}
    \toprule
    UI class                                      & Authorised & Anomaly & Uncertain \\
    \midrule
    Splash / OS startup                           & 0          & 2       & 0         \\
    Launcher / home                               & 6          & 1       & 0         \\
    Main flashlight control panel                 & 18         & 7       & 0         \\
    Settings \& theme-selection dialogs           & 2          & 10      & 0         \\
    Crash dialogs / \emph{App info} / blank state & 4          & 7       & 1         \\
    About screens                                 & 3          & 0       & 0         \\
    \midrule
    Total                                         & 33         & 27      & 1         \\
    \bottomrule
  \end{tabular}
\end{table}

\subsection{Summary of Findings}
RQ1 – Benign alignment / false positives. On the benign Wikipedia app, screens with high network/CPU usage that are clearly justified by the UI (e.g., voice search dialogues, loading spinners) are consistently classified as authorised. At the same time, anomalies concentrate on visually idle screens (“No saved pages yet”, empty Explore feed) that exhibit sustained memory/CPU plateaus. This suggests that \uixpose{} does not systematically flag legitimate heavy usage as anomalies, thereby mitigating false positives along typical interaction paths.

RQ2 – Malicious localisation / false negatives. For SmartCurrencyConverter, a known-malicious app, \uixpose{} surfaces clusters of anomalies in visually simple or error-state screens (chart failures, recommendation dialogues) where resource and network usage spike despite an apparently idle UI, aligning with reports of covert ad/analytics activity. This indicates that low-alignment, high-magnitude states can localise suspicious behaviour paths, reducing false negatives in these case studies.

RQ3 – Benign-but-buggy vs. malicious. In the benign Flashy app, anomalies concentrate around settings, theme dialogues, crash/App-info screens, and brief bursts after crashes. In contrast, the core control panel mostly shows modest, well-aligned behaviour. This pattern is consistent with inefficient resource management rather than hidden exfiltration, suggesting that the distribution of anomalies across UI classes can help distinguish buggy-but-benign apps from malicious ones.

%% file: sections/discussion.tex
\section{Discussion}\label{sec:discussion}

% The preliminary results demonstrate the promise of \uixpose{} and its novel design, while also revealing limitations in current approaches that highlight the need for further research and advancement.

The preliminary results in Section \ref{sec:evaluation} illustrate both the promise of \uixpose{} that is surfacing misaligned states in benign, malicious, and buggy apps, and several limitations of our current prototype. We now discuss these limitations and outline the research challenges they reveal.

\subsection{Lack of Direct UI-Automated Testing}

While \uixpose{} benefits from its modular and interdependent components, including UI automation and summarisation, a key limitation is the lack of a \textit{directed} UI testing mechanism, preliminary experiments indicate that existing tools, such as DroidBot, may spend significant time exploring uninteresting UI states that do not trigger network activity, for instance, repeatedly testing styling changes in the Wikipedia app. This highlights the need for targeted exploration strategies that prioritise states likely to produce observable behaviours, particularly network communications. Additionally, \uixpose{} requires UI automation tools compatible with non-native apps, such as those built with Unity, to enable interaction and exploration of a broader range of UI states.

\subsection{Lack of HTTP/S Payload Analysis}

Many existing studies overlook HTTP/HTTPS payloads in behavioural analysis, despite their potential to reveal critical anomalies. HTTPS traffic, in particular, is often ignored due to the need for decryption and specialised tools for interception. Integrating payload-level analysis alongside packet-level features would enhance the robustness of IBA analysis. \uixpose{} already captures and stores HTTPS payloads, but deeper integration is essential to leverage this data for fine-grained behavioural insights fully.

\subsection{Lack of Apps UI-Resource Utilisation}
Most Android malware research and benchmarks focus on static features, such as permissions and API calls, with few utilising detailed runtime resource data, including per-process CPU logs. Static methods, such as DREBIN \cite{arp2014drebin} and MaMaDroid \cite{onwuzurike2019mamadroid}, avoid on-device runtime monitoring due to resource constraints, so process-level resource utilisation data is not shared. Datasets mainly collect APKs or network flows, lacking host telemetry. Early anomaly research considered device-level CPU data, but neither standardised per-process traces nor created community datasets. Thus, there is no standard dataset linking malware samples with detailed resource usage, limiting the evaluation of detectors based on process dynamics. \uixpose{} already captures process-level resource utilisation, but building any AI-based model on such telemetry data is an essential step to obtain fine-grained behavioural insights.

\subsection{Limitations in UI Summarisation Across Domains}

Although \uixpose{} employs UI screen summarisation and intention inference to align with actual behaviours, it struggles with complex, dynamic UIs, even when previously analysed screens are available. Content from multiple endpoints and advertisements (e.g., UI elements in news or sports apps) hinders coherent summaries, obscuring the app's intent. More advanced summarisation techniques are needed to produce concise, user-centric descriptions of static and dynamic content~\cite{wang2021screen2words}. Without these, alignment with backend behaviour remains incomplete, limiting interpretability and effectiveness.

% \subsubsection{Scope of comparison.}

%% file: sections/conclusion.tex
\section{Conclusion and Future Work}
\label{sec:conclusion}

This paper presented the \uixpose{} framework for intention–behaviour alignment with
runtime semantics for mobile malware detection. \uixpose{} analyses application behaviour
via automated UI testing, summarisation, network payload analysis, and resource usage to
identify malicious activities that occur without user consent. It is a modular black-box
analysis framework that employs a co-instrumented runtime with an in-line network
monitor, logging UI callbacks, decoded traffic, and resource usage in a sandbox
environment. An alignment-guided exploration engine that integrates advanced automated
graphical user interface testing enables automated exploration of app screens through
generated user interactions. A screen-level intent module (vision–language with a
lightweight ontology) and a behaviour summariser over execution traces project UI intent
and observed behaviour into a shared action space, computing a quantitative misalignment
score that flags suspicious activity. Evaluation of three real-world cases showed that
\uixpose{} detects anomalies in communication, resource patterns, and payload semantics,
while highlighting practical challenges such as limited UI coverage, gaps in HTTPS
payload analysis, and the need for better UI summarisation tools. 

Recent LLM-based GUI testing frameworks, e.g., \cite{liu2025llm,wang2025llmdroid}, focus on
improving functional testing metrics (e.g., activity coverage, task completion, bug
finding) by using LLMs as UI drivers. These systems are therefore not directly comparable
to \uixpose{}, whose primary goal is to provide a security-centric oracle based on
intention–behaviour alignment. As future work, we plan to compare \uixpose{} against
intent-aware misbehaviour detectors, e.g., \cite{yang2022describectx,yoon2024intent}, and
to explore tighter integration with LLM-based GUI agents that use our alignment score as
a security signal within their exploration loops.

\section{Data Availability}

We provide the prototype implementation of \uixpose{} at: \url{https://github.com/amrmp/UIXpose}.